\documentclass[traditabstract,usenatbib]{aa}
\usepackage[varg]{txfonts}

\usepackage{mathtools}
\usepackage{aas_macros}
\usepackage{graphicx}
\usepackage{amstext}
\usepackage{enumitem}
\usepackage{amssymb}
\usepackage{booktabs}

\begin{document}

\newcommand{\mathleft}{\@fleqntrue\@mathmargin0pt}
\newcommand{\mathcenter}{\@fleqnfalse}
\newcommand{\unit}[1]{\ensuremath{\, \mathrm{#1}}}
\providecommand{\e}[1]{\ensuremath{\times 10^{#1}}}
\providecommand*{\printsecond}[2]{#2}
\newcommand{\edited}{}

\title{Significant uncertainties from calibrating overshooting with eclipsing binary systems}

\titlerunning{Uncertainties in DLEB overshooting constraints}

\author{Thomas Constantino \inst{\ref{inst1}}
        \and
        Isabelle Baraffe \inst{\ref{inst1},\ref{inst2}}
        }

  \institute{Physics and Astronomy, University of Exeter, Exeter, EX4 4QL, United Kingdom, \email{T.Constantino@exeter.ac.uk} \label{inst1} 
      \and
      Ecole Normale Sup\'erieure de Lyon, CRAL, UMR CNRS 5574, 69364 Lyon Cedex 07, France \label{inst2} 
      }

\authorrunning{Constantino and Baraffe}

\abstract{The precise measurement of the masses and radii of stars in eclipsing binary systems provides a window into uncertain processes in stellar evolution, especially mixing at convective boundaries.  Recently, these data have been used to calibrate models of convective overshooting in the cores of main sequence stars.  In this study we have used a small representative sample of eclipsing binary stars with $1.25 \leq M/\text{M}_\odot < 4.2$ to test how precisely this method can constrain the overshooting and whether the data support a universal stellar mass--overshooting relation.  We do not recover the previously reported stellar mass dependence for the extent of overshooting and in each case we find there is a substantial amount of uncertainty, that is, the same binary pair can be matched by models with different amounts of overshooting.  Models with a moderate overshooting parameter $0.013 \leq f_\text{os} \leq 0.014$ (using the scheme from \citealt{1997A&A...324L..81H}) are consistent with all eight systems studied.  Generally, a much larger range of $f_\text{os}$ is suitable for individual systems.  In the case of main sequence and early post-main sequence stars, large changes in the amount of overshooting have little effect on the radius and effective temperature, and therefore the method is of extremely limited utility.}


\maketitle

\section{Introduction}
\label{sec:introduction}

The treatment of mixing at convective boundaries is a fundamental uncertainty for stellar evolution calculations.  Basic arguments imply there must be some mixing beyond locally-determined convective boundaries according to, for example, the Schwarzschild criterion.  Theoretical estimates for the extent of overshooting vary considerably, ranging from very little to a zone of complete mixing around two pressure scale heights in depth.  The amount of overshooting in convective cores affects the main sequence lifetime and therefore the inferred age of stellar clusters and individual post-main sequence stars.  Convective core overshooting also increases the luminosity and speed of evolution of post-main sequence stars.

Several independent lines of evidence -- colour-magnitude diagrams of star clusters, double-lined eclipsing binary (DLEB) stars, and asteroseismology -- strongly suggest there is mixing beyond the Schwarzschild boundary of convective cores in main sequence stars.  By increasing the availability of hydrogen that can be burnt in the convective core, this mixing significantly extends the predicted main sequence lifetime.  There is currently no universally accepted theoretical basis to predict the extent of such mixing: it is typically dependent on a parameter (with or without a physical model).  In subsequent phases of evolution, the mixing beyond the Schwarzschild boundary of convection zones is equally crucial, but the relative scarcity of observational constraints means that the evolution is even more uncertain.  Characterizing and quantifying the processes operating in main sequence convective cores and convective boundaries may also help improve the models of later phases of stellar evolution.

Historically, several authors have proposed extensions to mixing length theory in order to quantify the amount of overshooting.  \citet{1965MNRAS.130..223R} argued that convective core overshooting region is of the order $10^{-3}$ times the stellar radius, which is up to about ten per cent of the radius of the convective core.  \citet{1973ApJ...184..191S} determined an average extent of convective overshoot of $0.01\,\text{M}_\odot$.  Adding more sophistication to the approach of \citet{1973ApJ...184..191S}, by accounting for the convective flux carried by overshooting elements and the resultant effect on the temperature gradient, \citet{1975ApJ...201..637C} arrived at 0.23 pressure scale heights of core overshooting for a 3$\,M_\odot$ star, in line with empirical estimates.  The applicability of these methods to stellar evolution calculations is limited by our lack of knowledge about the properties of convection in stellar cores and the difficulty of relating these penetration arguments to chemical mixing.

The best constraints for core overshooting so far have an empirical basis.  The most common approach has been to compare the width of the main sequence and shape of the turnoff observed colour-magnitude diagrams of stellar clusters with theoretical predictions.  \citet{1992A&AS...96..269S}, for example, {found that models with initial masses $1.25 \leq M/\text{M}_\odot \leq 25$ with 0.2\,pressure scale heights of overshooting were the best fitting for 65 observed clusters}.  Other studies have concluded that a similar magnitude of overshooting is needed.  This amount is often used as a default in stellar evolution codes and in published isochrones that are used widely by the astrophysical community.

The rationale for using DLEB stars to constrain main sequence overshooting is exactly the same as it is for using stellar clusters: they comprise stars born at the same time and with the same composition but different mass.  Although each system offers only a very limited insight compared with an entire stellar cluster, there is compensation from the high measurement precision.

\section{Evidence for mass-dependent overshooting}
\label{sec:mass_overshoot}

The simultaneous measurements of stellar mass $M$, radius $R$, and effective temperature $T_\text{eff}$ for presumably coeval and (initially) chemically identical DLEB stars has been used to investigate whether, and how, the amount of overshooting depends on stellar mass.  In recent years, there have been conflicting findings about the existence of such a trend.

\citet{1997MNRAS.285..696S} analysed nine DLEB pairs with stellar mass $2 \leq M/\text{M}_\odot \leq 7.2$ and found that the amount of required overshooting increases slightly with mass, from about $0.24\,H_\text{p}$ to $0.32\,H_\text{p}$ over the mass range examined.  \citet{1997MNRAS.289..869P} provided reason for caution about the potential of using DLEB stars to constrain the amount overshooting.  They found that 37 in their sample of 49 systems (nearly all from \citealt{1991A&ARv...3...91A}), the vast majority, could be satisfactorily matched by models both with and without overshooting.  They were able to do this by varying only the metallicity.

From a study of six DLEB pairs and three other non-eclipsing binary pairs in the stellar mass range $2 \lesssim M/\text{M}_\odot \lesssim 12$, \citet{2000MNRAS.318L..55R} reported that the amount of overshooting needs to increase with mass.  \citet{2007A&A...475.1019C} determined that overshooting was required in models of all ten stars in a sample with mass $M > 4\,\text{M}_\odot$.  Below that mass, nine of the 16 stars could be modelled without including overshooting.  These results were consistent with a small or non-existent mass dependence for overshooting above 2\,$\text{M}_\odot$.  In both these studies, the lack of data points prevented the detection of any mass-dependence below $2\,\text{M}_\odot$.

Recently, \citet{2016A&A...592A..15C} modelled 33 DLEB systems and found a mass dependence for the extent of overshooting.  They report that overshooting increases with stellar mass up to about $2\,\text{M}_\odot$ and then remains approximately constant.  In a further study, \citet{2017ApJ...849...18C} found the same trend using a different overshooting prescription.  Very recently, \citet{2018ApJ...859..100C} have analysed nine binary systems that have component(s) with stellar mass below 2\,M$_\odot$. The amount of overshooting required is consistent with earlier findings, specifically that there is a sharp increase between about 1.2 and 2\,M$_\odot$.  These studies stand in contrast with that of \citet{2015A&A...575A.117S} who used 11 DLEB pairs to find a large spread and no clear trend in the overshooting.

\citet{2017A&A...608A..62H} studied 19 systems with a very wide range of stellar masses, with primaries between $0.69\,\text{M}_\odot$ and $14.5\,\text{M}_\odot$.  They note the difficulty of constraining convective overshoot with their sample dominated by main sequence stars.  They find overshooting to be absolutely necessary in only two of the 14 cases where convective cores are present.  Overshooting is, however, favoured in a further six cases.  It is not clear how strongly this study supports the findings of \citet{2016A&A...592A..15C,2017ApJ...849...18C,2018ApJ...859..100C}.  Among the low mass pairs, \citet{2017A&A...608A..62H} find AI PHI (1.23\,M$_\odot$ and 1.19\,M$_\odot$) models required no overshooting, UXMEN (1.24\,M$_\odot$ and 1.20\,M$_\odot$) and KOI-3571 (1.24\,M$_\odot$ and 1.09\,M$_\odot$) give similar results with and without overshoot, V501 Her (1.27\,M$_\odot$ and 1.21\,M$_\odot$) and KIC 9777062 (1.60\,M$_\odot$ and 1.42\,M$_\odot$) are better fitted with overshooting, and overshooting is required for BG Ind (1.43\,M$_\odot$ and 1.29\,M$_\odot$).  In this mass range, overshooting is indeed more favoured for greater stellar mass, but the results are less conclusive than \citet{2016A&A...592A..15C,2017ApJ...849...18C,2018ApJ...859..100C}.

\citet{2016A&A...587A..16V} analysed theoretical uncertainties for helium content, metallicity, $T_\text{eff}$, mass, radius, MLT mixing length, and element diffusion for systems with stellar masses $1.1 \leq M/\text{M}_\odot \leq 1.6$.  They conclude that this method for establishing the extent of overshoot is unreliable, especially for stars yet to reach the end of the main sequence.  Later, \citet{2018arXiv180307058V} tested the sensitivity of these methods to typical uncertainties for an evolved system containing a $2.50\,\text{M}_\odot$ primary and a $2.38\,\text{M}_\odot$ secondary.  They find a systematic uncertainty in the amount of overshooting of $\pm 20$ per cent for stars evolved beyond the main sequence, and in some cases systematic biases (such as higher overshooting when the true overshooting is small).  The situation is even worse for stars near the end of the main sequence.  The lack of consensus in these recent studies of main sequence overshooting--a vital factor in stellar evolution calculations--demands we give the subject further attention.

\section{Stellar models}
\label{sec:models}

In this paper we calculate the stellar evolution sequences using the Monash/Mt Stromlo code {\sc monstar} \citep[see e.g.][]{2008A&A...490..769C}.  The hydrogen- and helium-burning reaction rates are from \citet{1999NuPhA.656....3A}.  The low-temperature ($T < 10000\,\text{K}$) opacity tables were generated using the AESOPUS tool (\citealt{2009A&A...508.1539M}; see \citealt{2014ApJ...784...56C} for details of the implementation).  The high-temperature opacity tables are from \citet{1996ApJ...464..943I}.  In this study we use the 2005 update to the OPAL equation of state \citep{2002ApJ...576.1064R} except in the high-temperature and high-density regimes where the Helmholtz equation of state \citep{2000ApJS..126..501T} is used.

In each eclipsing binary case we explore models with a range of metallicity (with the heavy element abundances scaled according to the \citealt{2009ARA&A..47..481A} solar determination; hereafter A09) and MLT mixing length $\alpha_\text{MLT}$, if relevant.  We aim to establish whether the solution for the extent of overshooting is unique for each system, or if there are a range of model solution with reasonable assumptions that are consistent with the observations and their uncertainties.  All but one binary pair comprise two stars of very similar (and all except two nearly identical) mass, which may minimize the impact of other uncertainties inherent in stellar models.  {We consider a pair of models to be valid solution if each member matches the observed radius and $T_\text{eff}$ to within the uncertainty reported in the literature.}  {We require solutions in which the two members of each system have an identical initial composition and MLT parameter, which is generally well justified due to their similar stellar parameters ($M$, $T_\text{eff}$, $\log{g}$)}.  In each case, we begin our search with $\alpha_\text{MLT} = 1.60$, which is the {\sc monstar} solar calibrated value.  The models have initial helium abundance $Y=0.25$ or $Y=0.26$, depending on the expected metallicity\footnote{While the latter is slightly lower than commonly adopted for solar-metallicity models, our tests show the conclusions are not affected.}.

We generically refer to mixing beyond the Schwarzschild boundary as `overshoot' without implying any particular mechanism.  We model this mixing using the widely adopted scheme proposed by \citet{1997A&A...324L..81H} based on the 2D hydrodynamical simulations from \citet{1996A&A...313..497F}, where the diffusion coefficient near convective boundaries $D_\text{os}$ is given by
\begin{equation}
D_\text{os}(z) = D_0 e^{-2z/f_\text{os}H_\text{p}},
\label{eq_herwig}
\end{equation}
where $D_0$ is the diffusion coefficient inside the convective boundary derived from MLT, $z$ is the distance from the boundary, $H_\text{p}$ is the pressure scale height at the convective boundary, and $f_\text{OS}$ is a free parameter. \citet{2017ApJ...849...18C} compared the overshooting trend calculated with this prescription to that from using step-overshooting (that is complete mixing over a certain distance) and demonstrated that the trend was essentially independent of the scheme, which is consistent with the basic picture that the important factor in the evolution is the mass enclosed by the well-mixed region.

We adopt only the \citet{1997A&A...324L..81H} prescription to allow for mixing in formally convectively stable regions.  This therefore acts as a proxy for any other process which has the effect of mixing material near the convective core.  Mixing resulting from rotation has been invoked to explain the extended main sequence turnoffs observed in stellar clusters \citep[e.g.][]{2009MNRAS.398L..11B}.  This is because rotation may cause chemical mixing in convectively stable regions and also affect the observed colour and brightness because they depend on the orientation of the rotation axis \citep[see e.g.][]{2011A&A...533A..43E}.  These effects immediately highlight two problems for using eclipsing binaries to constrain the extent of overshoot: (i) there may be a degeneracy between the extent of the convective core and rotationally induced mixing and (ii) the magnitude and temperature inferred may depend on the orientation of each star.

The specific value for the required overshooting parameter will depend on how the mixing scheme is implemented.  When the scheme according to \citet{1997A&A...324L..81H} is used, the amount of mixing depends on where inside the convection zone the exponential decrease in the diffusion coefficient begins (this cannot be at the convective boundary because in MLT the convective velocity vanishes there).  The results will similarly depend on microphysics (such as equation of state and opacity) as well as the composition adopted.  We do not expect, however, that the existence of any overall trend between mass and overshooting would be affected.

The choice of $\alpha_\text{MLT}$, which is poorly constrained other than for near-surface convection also affects the implied diffusion coefficient in the convection zone (it increases with increasing $\alpha_\text{MLT}$) and hence also in the overshooting region.  The mixing, however, is considerably more sensitive to changes in $f_\text{os}$ than $\alpha_\text{MLT}$.  In the case of small convection zones, the standard choice for $\alpha_\text{MLT}$ may imply that the local mixing length $l = \alpha_\text{MLT} H_\text{p}$ is greater than the depth of the convection zone, perhaps giving an unrealistically high estimate of the diffusion coefficient.

In the models with step-overshooting in the literature, the amount of overshooting may be described by a parameter $\alpha_\text{os}$, but this can have different meanings: it is either the overshooting length expressed in units of the pressure scale height $H_\text{P}$ at the Schwarzschild boundary of the convective core or expressed in units of the radius of the convective core $r_\text{cc}$.  This inconsistency can make comparing results difficult.  \citet{2016A&A...592A..15C}, for example calculate the overshooting length $l$ using
\begin{equation}
l = 
\begin{cases} 
 \alpha_\text{os} H_\text{P} & \text{if $r_\text{cc} < H_\text{P}$} \\
 \alpha_\text{os} r_\text{cc} & \text{if $r_\text{cc} \ge H_\text{P}$} \\
\end{cases}
,
\end{equation}
whereas studies such as that from \citet{2015A&A...575A.117S} always report the overshooting length as a fraction of $H_\text{P}$.
Below about $2\,\text{M}_\odot$ the convective core radius happens to be around 1\,$H_\text{P}$ so the effect of this choice is small there, but at 4\,$\text{M}_\odot$ it is approximately $2\,H_\text{P}$.

\subsection{Selection of eclipsing binary systems}

In this paper we select a sub-sample of eight eclipsing binary systems from the sets previously analysed by \citet{2016A&A...592A..15C}, \citet{2017ApJ...849...18C}, or \citet{2018ApJ...859..100C}.  Our sub-sample was chosen to be representative of the stellar mass and evolutionary stage of those larger samples.  The properties of the stars in our sub-sample are presented in Table~\ref{table_results_summary}.

\section{Models of hydrogen burning stars}

\subsection{SZ Cen}
\label{sec:SZ_cen}
SZ Cen is a well studied system with a 2.311$\,\text{M}_\odot$ primary and a 2.272$\,\text{M}_\odot$ secondary \citep{2010A&ARv..18...67T}.  \citet{1975A&A....45..203A} and \citet{1977A&A....55..401G} were unable to find acceptable fits using models without convective overshoot.  \citet{1991A&ARv...3...91A} determined that the primary of SZ Cen must be in a rapid (post-main sequence) phase of evolution using models with overshooting from \citet{1988A&AS...76..411M,1989A&A...210..155M}.  \citet{1997MNRAS.289..869P} found models with and without overshoot to be equally good fits.  \citet{10.1007/978-94-011-4497-1_11} concluded that only models with overshooting \citep{1978A&A....65..281R,1989A&A...211..361R} could match the system.

SZ Cen has also been included in several more recent studies quantifying the required amount of overshoot \citep{2000MNRAS.318L..55R,2007A&A...475.1019C,2015A&A...575A.117S, 2016A&A...592A..15C,2017ApJ...849...18C}.  \citet{2000MNRAS.318L..55R} arrived at $ 0.1 \leq \alpha_\text{os} \leq 0.2$ and \citet{2007A&A...475.1019C} reported $0.0 \leq \alpha_\text{os} \leq 0.2$.  \citet{2015A&A...575A.117S} found $f_\text{os} = 0.025$, whereas \citet{2017ApJ...849...18C} determined $f_\text{os} = 0.0165$ and $f_\text{os} = 0.0195$ using scaled solar mixtures from \citet{1998SSRv...85..161G} and A09, respectively.

\begin{figure}
\includegraphics[width=\linewidth]{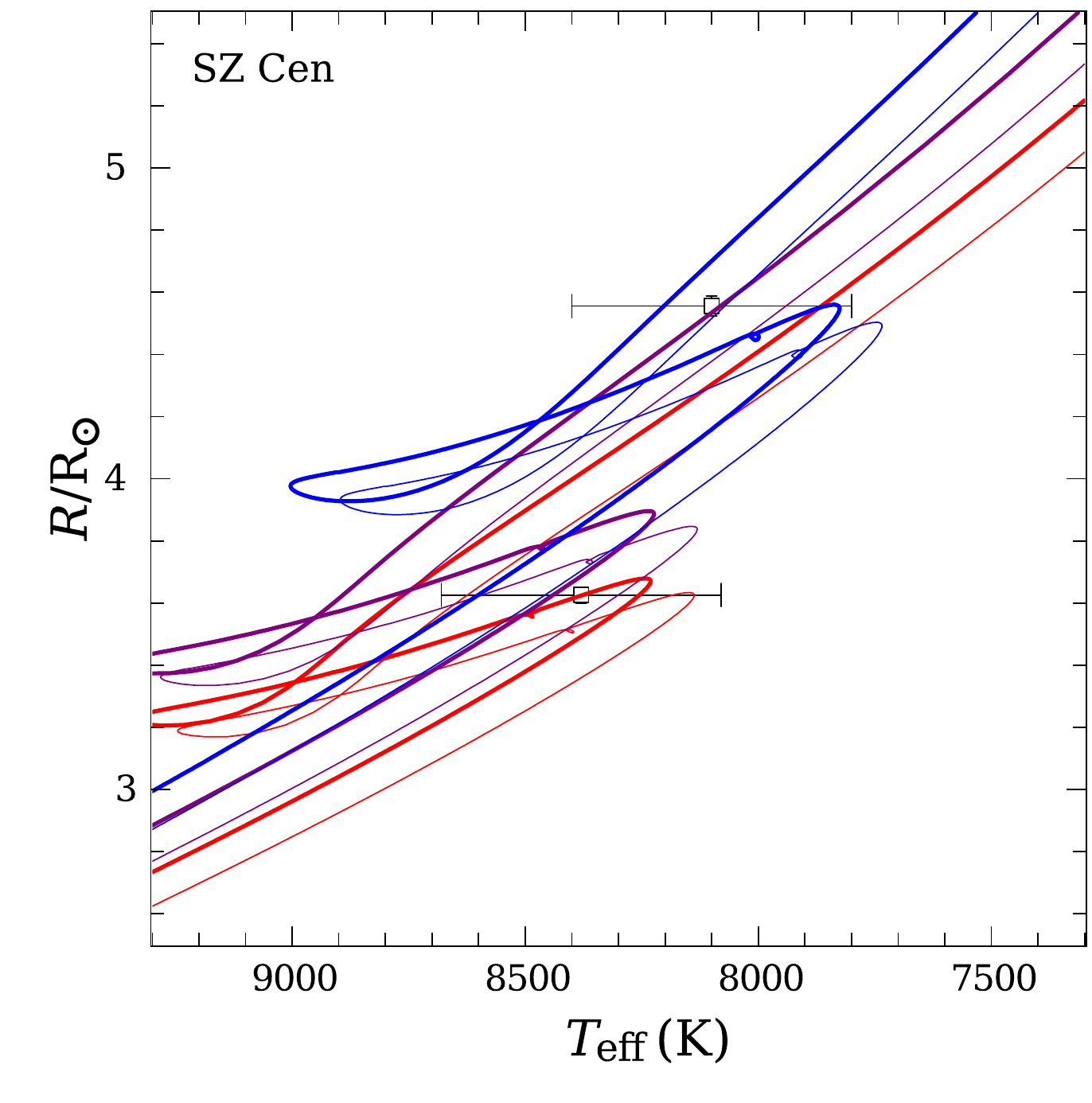}
  \caption{Evolution tracks of stellar models of SZ~Cen.  Thick lines denote the primary models.  Error bars indicate the uncertainty in $R$ and $T_\text{eff}$ reported by \citet{2010A&ARv..18...67T}.  The models have $0.013 \leq f_\text{os} \leq 0.028$ and $-0.25 \leq \text{[Fe/H]} \leq -0.20$.  The redder curves show models with lower $f_\text{os}$.}
  \label{figure_SZCen}
\end{figure}

Our models agree with the earlier consensus that some overshooting is required to match the components of the SZ Cen system.  We were able to construct satisfactory models with a range of overshooting $0.013 \leq f_\text{os} \leq 0.028$ within a narrow metallicity range $-0.25 \leq \text{[Fe/H]} \leq -0.20$ (the high-metallicity end of this range corresponds to $Z = 0.0090$, the best fit solution from \citealt{2017ApJ...849...18C}).  A selection of those models, with $f_\text{os} = 0.013$, $f_\text{os} = 0.018$, and $f_\text{os} = 0.028$ are presented in Figure~\ref{figure_SZCen}.  When the overshooting is below about $f_\text{os} = 0.028$, the solution has the primary having just finished convective core hydrogen burning and beginning to move towards the red giant branch during a relatively rapid phase of evolution.  If the overshooting is larger, there are solutions with both stars still on the main sequence, and because the evolution in $T_\text{eff} - R$ space is slower there, this appears to be a more favourable solution.

\begin{figure}
\includegraphics[width=\linewidth]{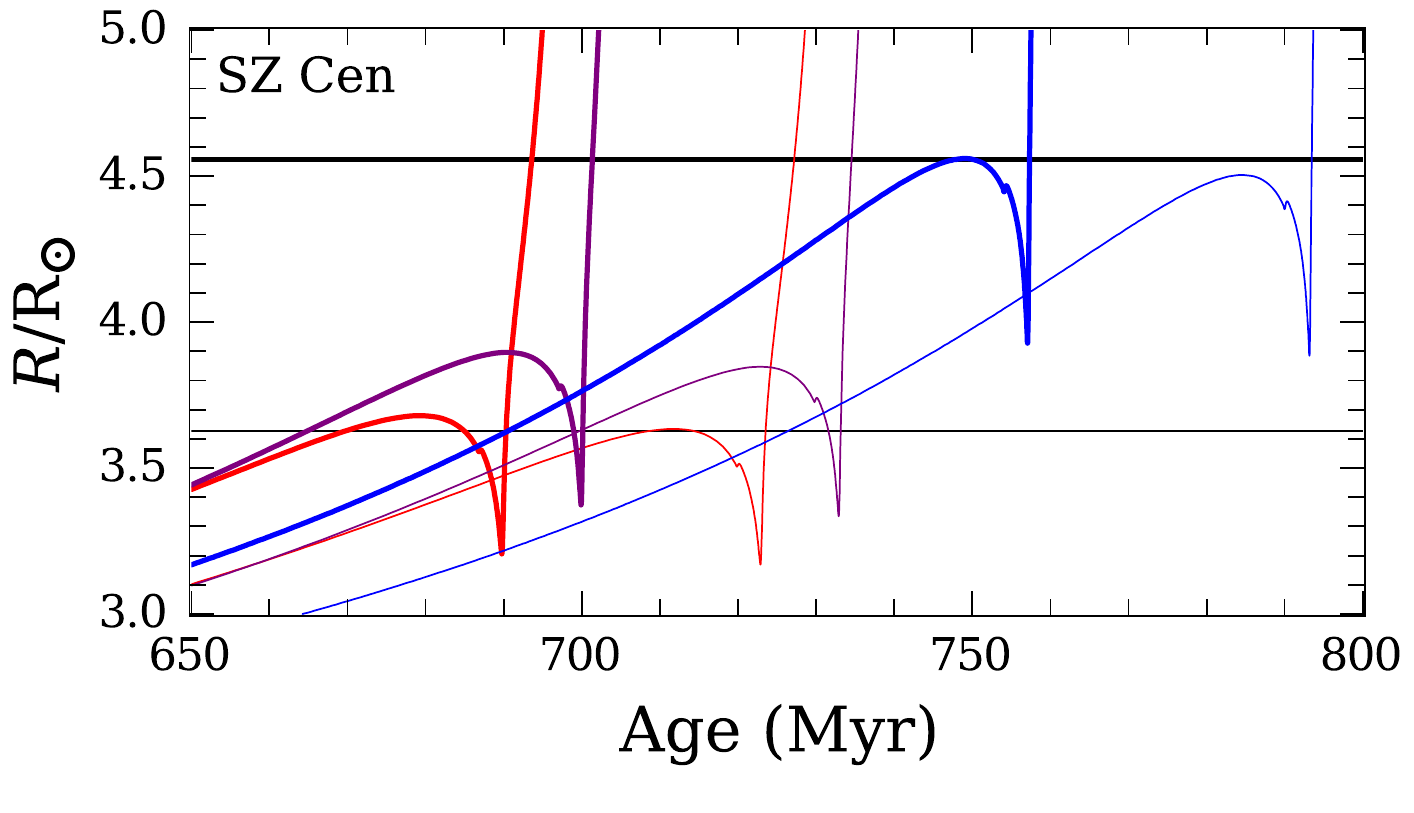}
  \caption{Evolution of radius of the SZ~Cen models shown in Figure~\ref{figure_SZCen}.  The colours are the same as Figure~\ref{figure_SZCen} and the thick and thin horizontal lines show the primary and secondary radius, respectively.}
  \label{figure_SZCen_R}
\end{figure}

\citet{1975A&A....45..203A} found a mass ratio $q= M_\text{primary}/M_\text{secondary} = 1.017 \pm 0.007$.  The uncertainty in this mass ratio corresponds to an age difference of around 14\,Myr.  In Figure~\ref{figure_SZCen_R} we show the radius evolution for the three pairs of models shown in Figure~\ref{figure_SZCen}.  In the pair with the lowest $f_\text{os}$, the correct $R$ and $T_\text{eff}$ are found when the primary is around 15\,Myr younger than the secondary.  This difference would be even worse for lower $f_\text{os}$ because the secondary would not attain the observed $R$ and $T_\text{eff}$ until after the end of the main sequence.  The pair with $f_\text{os} = 0.018$ have a nearly identical age.  In the pair with the highest $f_\text{os}$, the primary is about 20\,Myr older than its companion.  This is still less than three per cent of the age of the system and the majority of the discrepancy could be explained by the uncertainty in the mass ratio.  This age discrepancy is also less than the maximum five per cent allowed by \citet{2016A&A...592A..15C,2017ApJ...849...18C,2018ApJ...859..100C}.  We conclude that SZ Cen is moderately useful for constraining the extent of overshooting because we were able to establish there is a likely lower limit (from the implied from the age difference and the lower likelihood of both stars being in faster stages of evolution).


\subsection{AY Cam}
\label{sec:AY_Cam}

AY Cam has a 1.905$\,\text{M}_\odot$ primary and 1.709$\,\text{M}_\odot$ secondary \citep{2010A&ARv..18...67T}.  \citet{2015A&A...575A.117S} found a best fit of $f_\text{os} = 0.020$ for the primary and $f_\text{os} = 0.019$ for the secondary.  The small discrepancy results from their requirement of a smooth increase in overshooting between 1.1 and $1.8\,\text{M}_\odot$ \citep[cf.][]{2006ApJS..162..375V}.  \citet{2017ApJ...849...18C} found $f_\text{os} = 0.015$ for the primary and $f_\text{os} = 0.014$ for the secondary, respectively, using the A09 mixture.

\begin{figure}
\includegraphics[width=\linewidth]{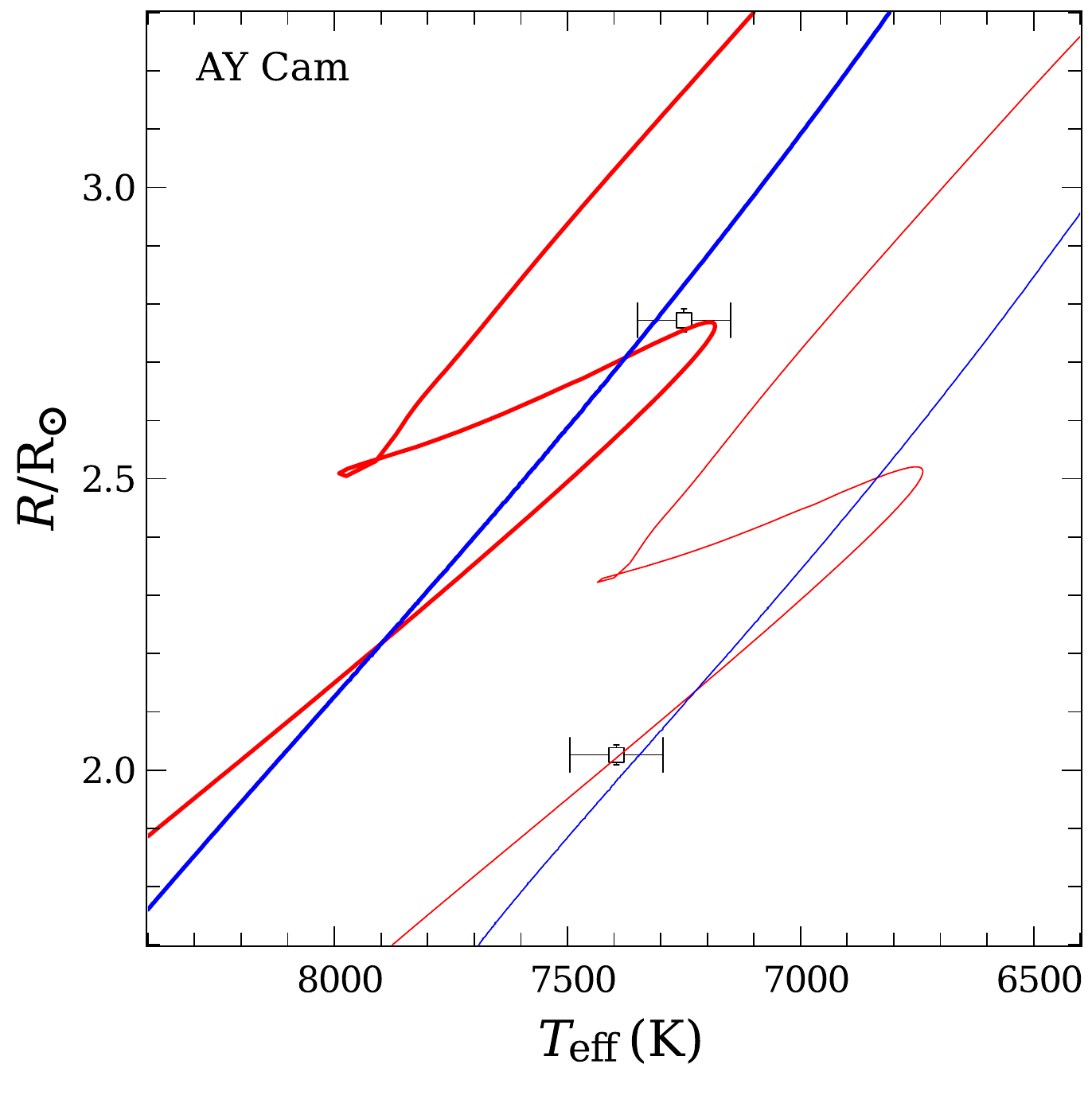}
  \caption{Evolution tracks of stellar models of AY~Cam.  Thick lines denote the primary models.  Error bars indicate the uncertainty in $R$ and $T_\text{eff}$ reported by \citet{2010A&ARv..18...67T}.  The models have $f_\text{os} = 0.0$ and $\text{[Fe/H]} = 0.081$ (in red), and $f_\text{os} = 0.04$ and $\text{[Fe/H]} = 0.180$ (in blue). }
  \label{figure_AYCam}
\end{figure}

\begin{figure}
\includegraphics[width=\linewidth]{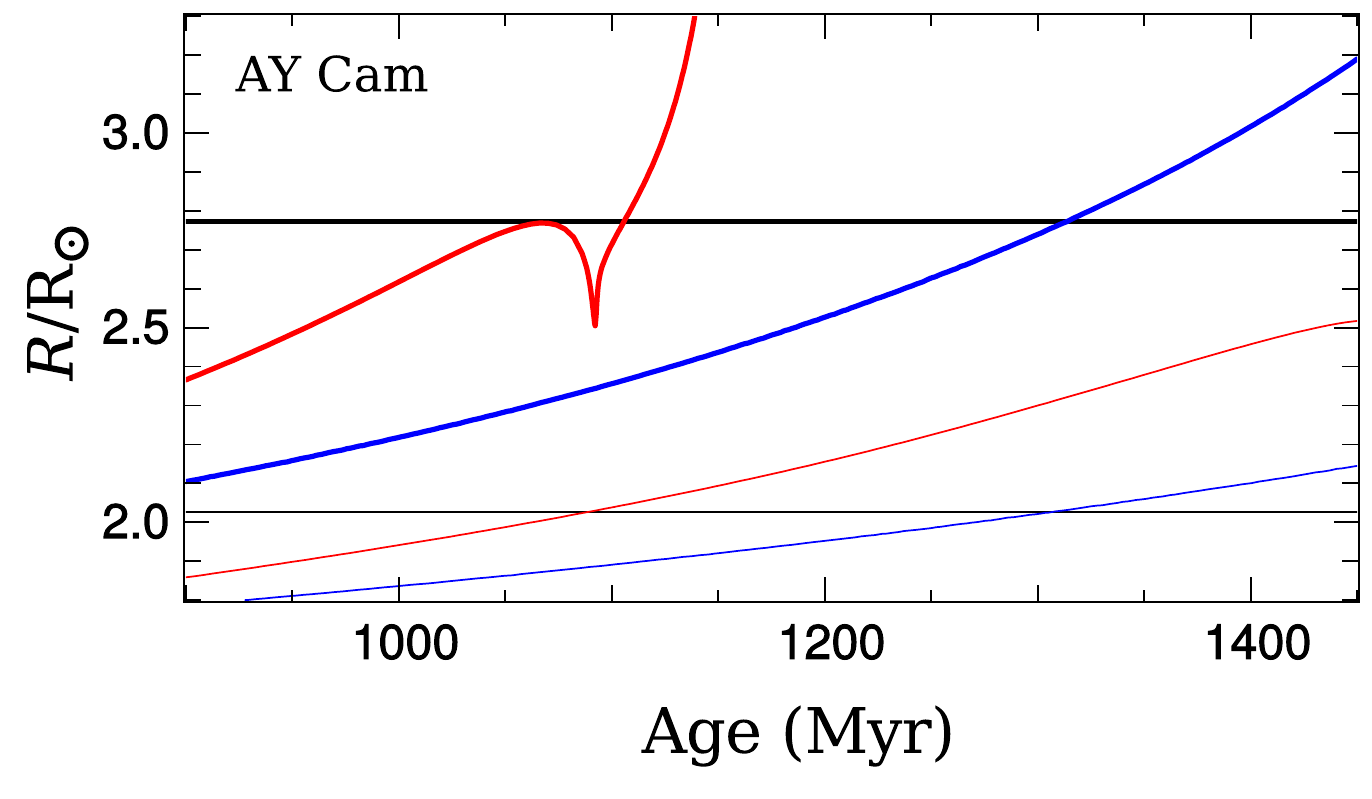}
  \caption{Evolution of radius of the AY~Cam models shown in Figure~\ref{figure_AYCam}.  The colours are the same as Figure~\ref{figure_AYCam} and the thick and thin horizontal lines show the primary and secondary radius, respectively.}
  \label{figure_AYCam_R}
\end{figure}

In our tests, we restricted our search to models with the same $f_\text{os}$ for the primary and secondary because of the similarity between their masses.  It is apparent from Figure~\ref{figure_AYCam} that the best fits will have both stars on the main sequence, that is before the primary temporarily moves to higher $T_\text{eff}$ near the end of core hydrogen burning.  We were able to produce satisfactory pairs of models with a large range of overshoot $0.00 \leq f_\text{os} \leq 0.04$ by making small increases in metallicity when increasing $f_\text{os}$ (from $\text{[Fe/H]} = 0.081$ to $0.180$).  The lower end of this metallicity range closely corresponds to the $Z = 0.0150$ best fit models from \citet{2017ApJ...849...18C}.  Figure~\ref{figure_AYCam_R} shows that an age difference between the two components emerges as $f_\text{os}$ increases, discrediting models with any higher $f_\text{os}$.  We conclude that AY Cam is of limited use for constraining overshooting because we found solutions with a broad range of $f_\text{os}$ by making only small adjustments to the metallicity.

\subsection{HD 187669}
\label{sec:HD187669}

\begin{figure}
\includegraphics[width=\linewidth]{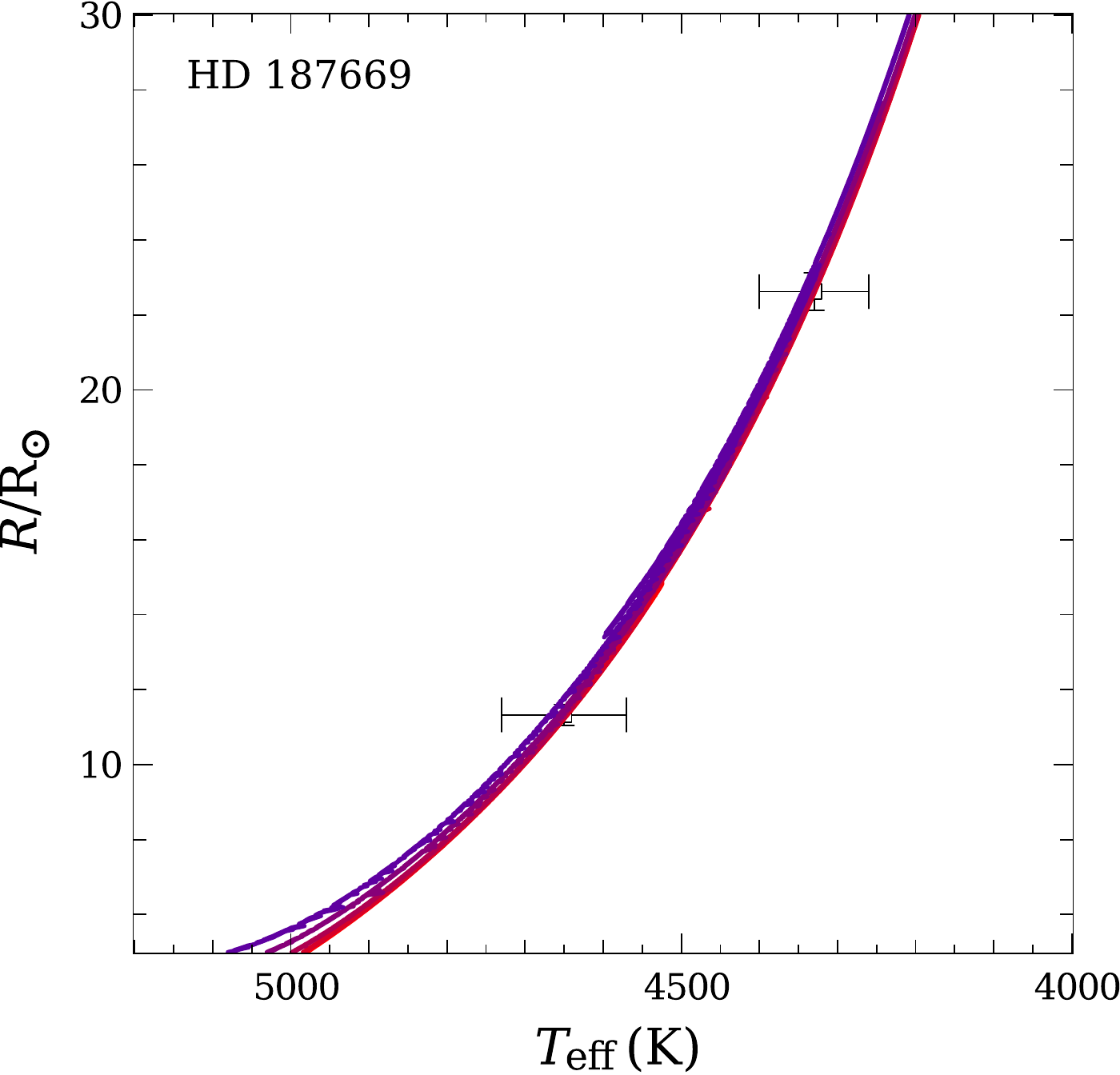}
  \caption{Evolution tracks of stellar models of HD~187669.  Thick curves denote the primary models.  Error bars indicate the uncertainty in $R$ and $T_\text{eff}$ reported by \citet{2015MNRAS.448.1945H}.  The models have metallicity $\text{[Fe/H]} = -0.25$ and $0.00 \leq f_\text{os} \leq 0.04$.}
  \label{figure_HD187669}
\end{figure}

\begin{figure}
\includegraphics[width=\linewidth]{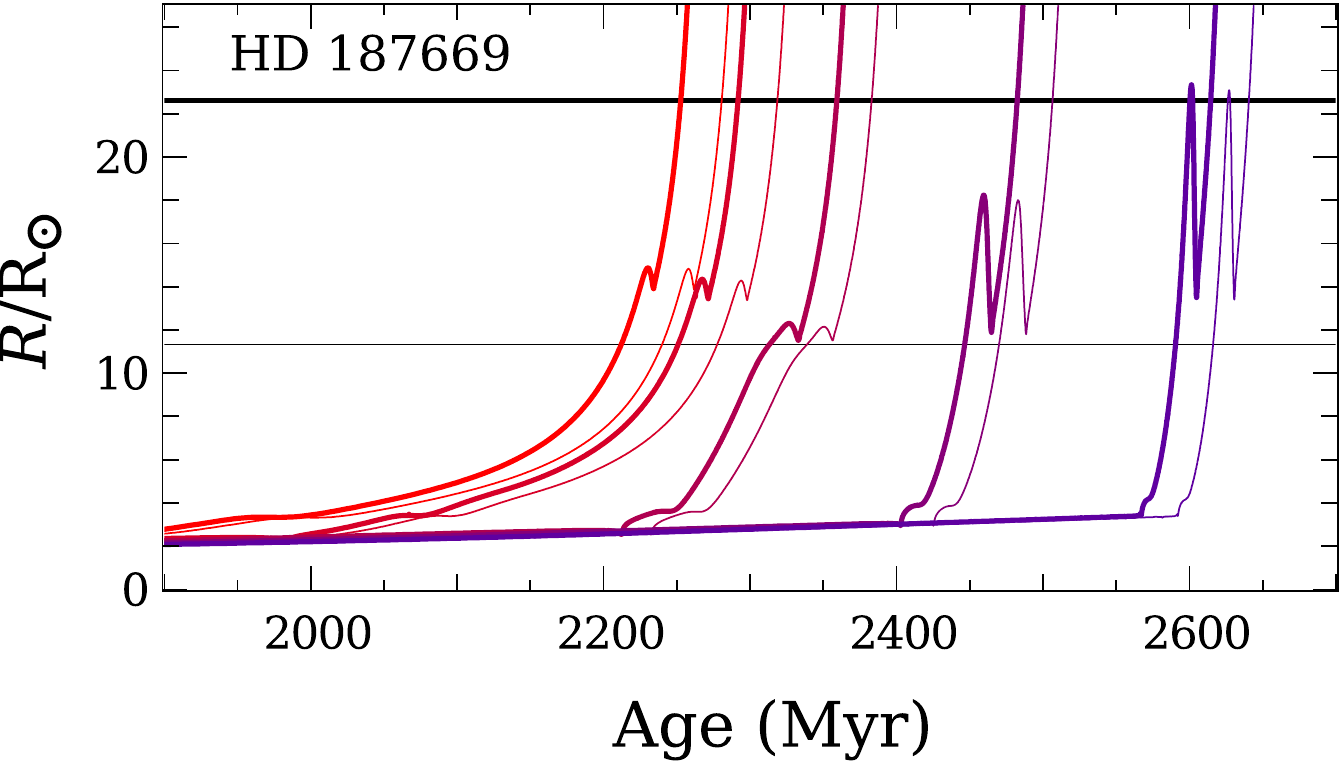}
  \caption{Evolution of radius of the HD~187669 models shown in Figure~\ref{figure_HD187669}.  The colours are the same as Figure~\ref{figure_HD187669} and the thick and thin horizontal lines show the primary and secondary radius, respectively.}
  \label{figure_HD187669_R}
\end{figure}

HD~187669 comprises two stars of nearly identical mass, a 1.505$\,\text{M}_\odot$ primary and a 1.504$\,\text{M}_\odot$ secondary \citep{2015MNRAS.448.1945H}.  \citet{2017ApJ...849...18C} found $f_\text{os} =0.009$ for both components of the HD~187669 system.  Rather than plotting two sets of models with essentially identical mass we present models of a secondary with $M = 1.500\,\text{M}_\odot$ in Figures~\ref{figure_HD187669} and \ref{figure_HD187669_R}.  Figure~\ref{figure_HD187669} shows that all of the evolution tracks computed with $0.00 \leq f_\text{os} \leq 0.04$ and $\text{[Fe/H]} = -0.25$ (the spectroscopic value arrived at by \citealt{2015MNRAS.448.1945H}) pass almost exactly through the observed data points in $T_\text{eff} - R$ space.  The small mass difference between components leads only to a difference in time taken to reach the same point in $ R - T_\text{eff}$ space.  In the worst case for solutions, where the two components have equal mass, the age difference between the two components is less than two per cent.  Figure~\ref{figure_HD187669_R} shows that with moderate overshooting ($f_\text{os} \approx 0.02$) the tracks of the secondary are close to passing through the observed position in the $T_\text{eff} - R$ diagram multiple times.  When the overshooting is large ($f_\text{os} \geq 0.04$) the primary passes the observed position in the $T_\text{eff} - R$ diagram three times.  

The best fit models from \citet{2015MNRAS.448.1945H} have the secondary as a post-main sequence star, whereas it is still burning hydrogen in a convective core in our simulations.  They also suggest the majority of the system's age uncertainty results from uncertainty in [Fe/H].  However, this ignores the effect of overshooting, which in these tests can change the age by 0.4\,Gyr without a change in metallicity.

We find that HD~187669 is not useful for constraining overshoot for two main reasons: (i) the amount of overshooting scarcely affects the path of the evolution in $T_\text{eff} - R$ diagram, and (ii) the age differences between our models are smaller than two per cent.


\subsection{$\chi^2$ Hya}
\label{sec:chi2Hya}

$\chi^2$ Hya is the system in this paper with the largest mass difference between components: it has a $3.605\,\text{M}_\odot$ primary and a $2.632\,\text{M}_\odot$ secondary \citep{2010A&ARv..18...67T}.  \citet{1997MNRAS.289..869P} found that models with and without overshoot were consistent with the observations.  \citet{2007A&A...475.1019C} came to a similar conclusion, finding $\alpha_\text{os} = 0.2^{+0.1}_{-0.2}$ for both members.  \cite{2014ApJ...787..127M} found a very large uncertainty in their calibration of the overshooting parameter in scheme from \citet{2013ApJS..205...18Z}.  \citet{2016A&A...592A..15C} found a best fit of  $\alpha_\text{os} = 0.200$ for each component but with an age difference greater than five per cent.

\begin{figure}
\includegraphics[width=\linewidth]{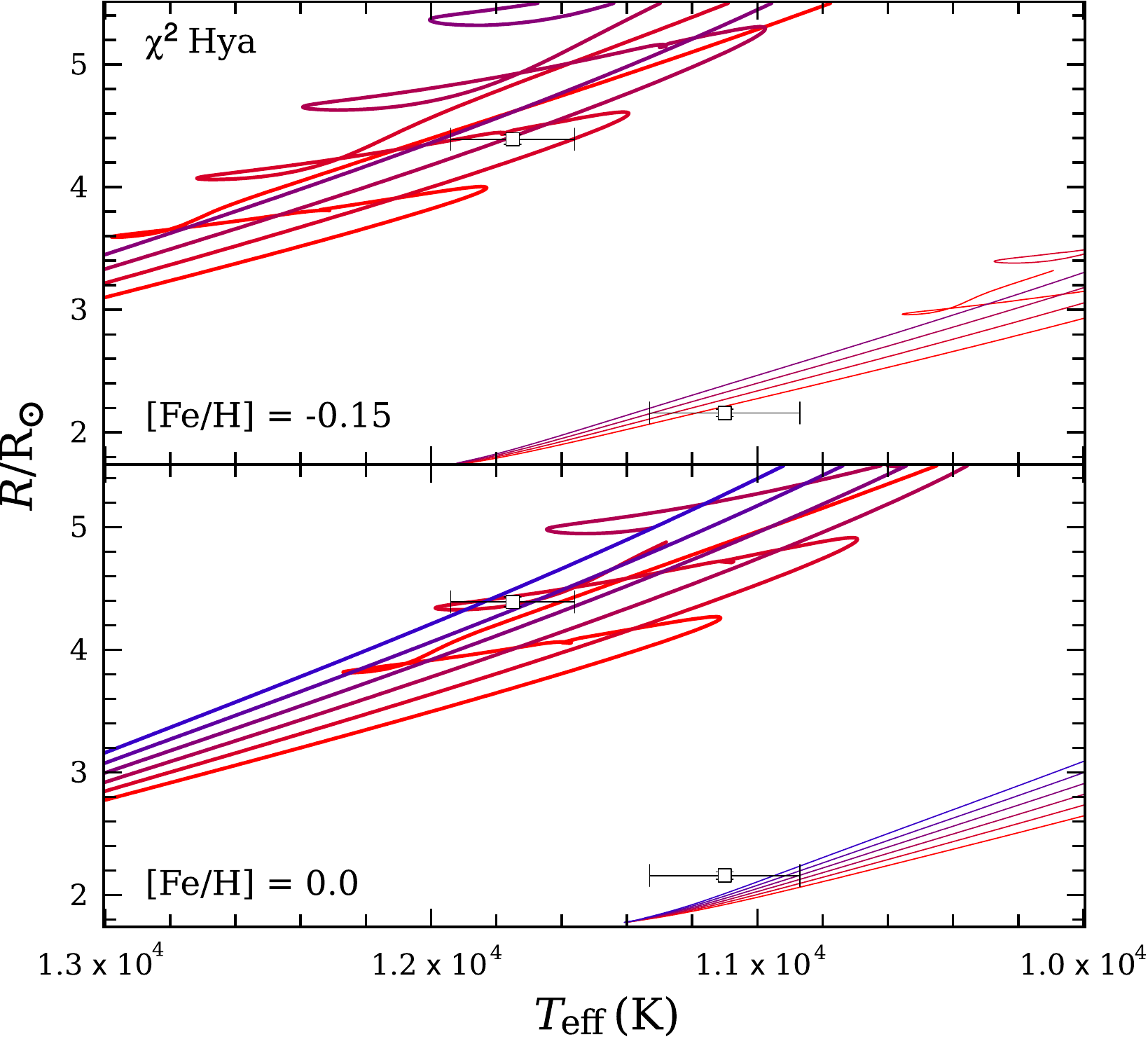}
  \caption{Evolution tracks of stellar models of $\chi^2$~Hya.    Thick lines denote the primary models.  Error bars indicate the uncertainty in $R$ and $T_\text{eff}$ reported by \citet{2010A&ARv..18...67T}.  The models in the upper panel have $\text{[Fe/H]} = -0.15$ and $0.00 \leq f_\text{os} \leq 0.03$ and those in the lower panel have $\text{[Fe/H]} = 0.0$ and $0.00 \leq f_\text{os} \leq 0.05$.}
  \label{figure_chi2_Hya}
\end{figure}

\begin{figure}
\includegraphics[width=\linewidth]{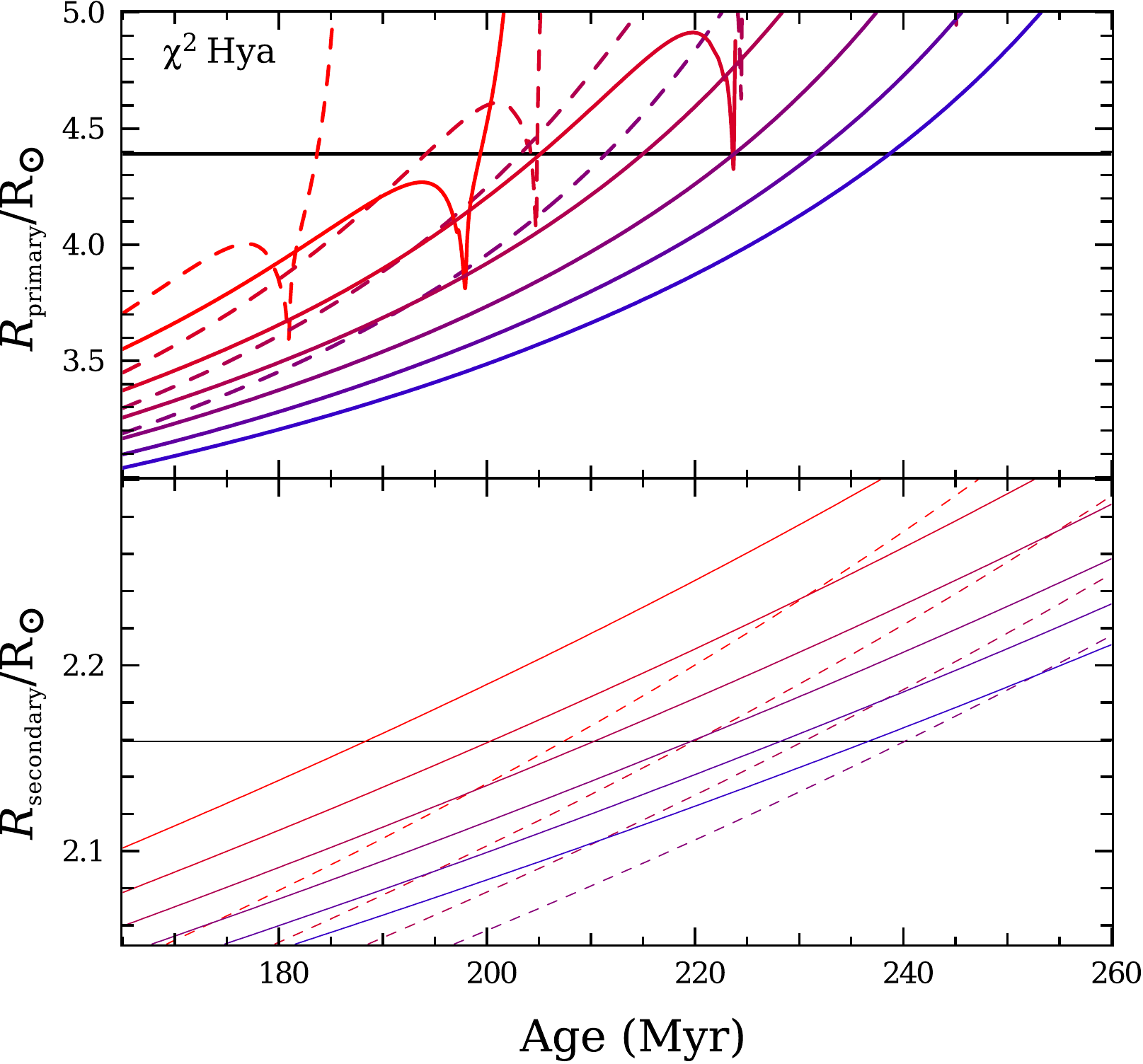}
  \caption{Evolution of radius of the $\chi^2$~Hya models shown in Figure~\ref{figure_chi2_Hya}.  The colours are the same as Figure~\ref{figure_chi2_Hya} and the thick and thin horizontal lines show the primary and secondary radius, respectively.}
  \label{figure_chi2_Hya_R}
\end{figure}

Figure~\ref{figure_chi2_Hya} shows our models for $\chi^2$ Hya with two metallicities $\text{[Fe/H]} = -0.15$ and $\text{[Fe/H]} = 0.0$, which are both around the $Z = 0.0110$ value used for the best fit from \citet{2016A&A...592A..15C}.  We find a large range of overshooting is consistent with the observations.  Models with little overshooting, $f_\text{os} \leq 0.02$, are favoured when the metallicity is low.  Figure~\ref{figure_chi2_Hya_R} shows that the secondaries have a significantly older predicted age than the primaries when $\text{[Fe/H]} = -0.15$ and $f_\text{os} = 0.0$.  However, when $\text{[Fe/H] = 0.0}$ the secondaries are younger than the primaries.  Together, these results imply that there are possible matches with consistent ages and $-0.15 < \text{[Fe/H]} < 0.0$ across the entire overshooting range $0.00 \leq f_\text{os} \leq 0.05$.  Adopting either of the metallicities tested, $\text{[Fe/H]} = -0.15$ or $\text{[Fe/H]} = 0.0$, the $f_\text{os} = 0.0$ primary is in the faster post-main sequence phase when it reaches the required radius, suggesting it is a lower probability fit.

\subsection{BK Peg}
\label{sec:BKPeg}

BK~Peg comprises the two lowest mass stars in this paper: a $1.414\,\text{M}_\odot$ primary and a $1.257\,\text{M}_\odot$ secondary.   \citet{2018ApJ...859..100C} found a best fit metallicity of $Z = 0.015$ using the A09 mixture, overshooting parameters $f_\text{os} = 0.008$ and $0.000$, and mixing length parameters $\alpha_\text{MLT} = 1.90$ and $2.03$ for the primary and the secondary, respectively.  The amount of overshooting required for the two components places them nicely on the fit shown in their Figure~2, where there is a roughly linear growth of $f_\text{os}$ between about $1.2\,\text{M}_\odot$ and $1.8\,\text{M}_\odot$.  The metallicity of those models is a little higher than implied by the spectroscopically determined value of $\text{[Fe/H]} = -0.12 \pm 0.07$ \citep{2010A&A...516A..42C}.  In their comparison with evolution tracks, \citet{2010A&A...516A..42C} also found a higher metallicity, $\text{[Fe/H]} = -0.05$, to be a better match.

\begin{figure}
\includegraphics[width=\linewidth]{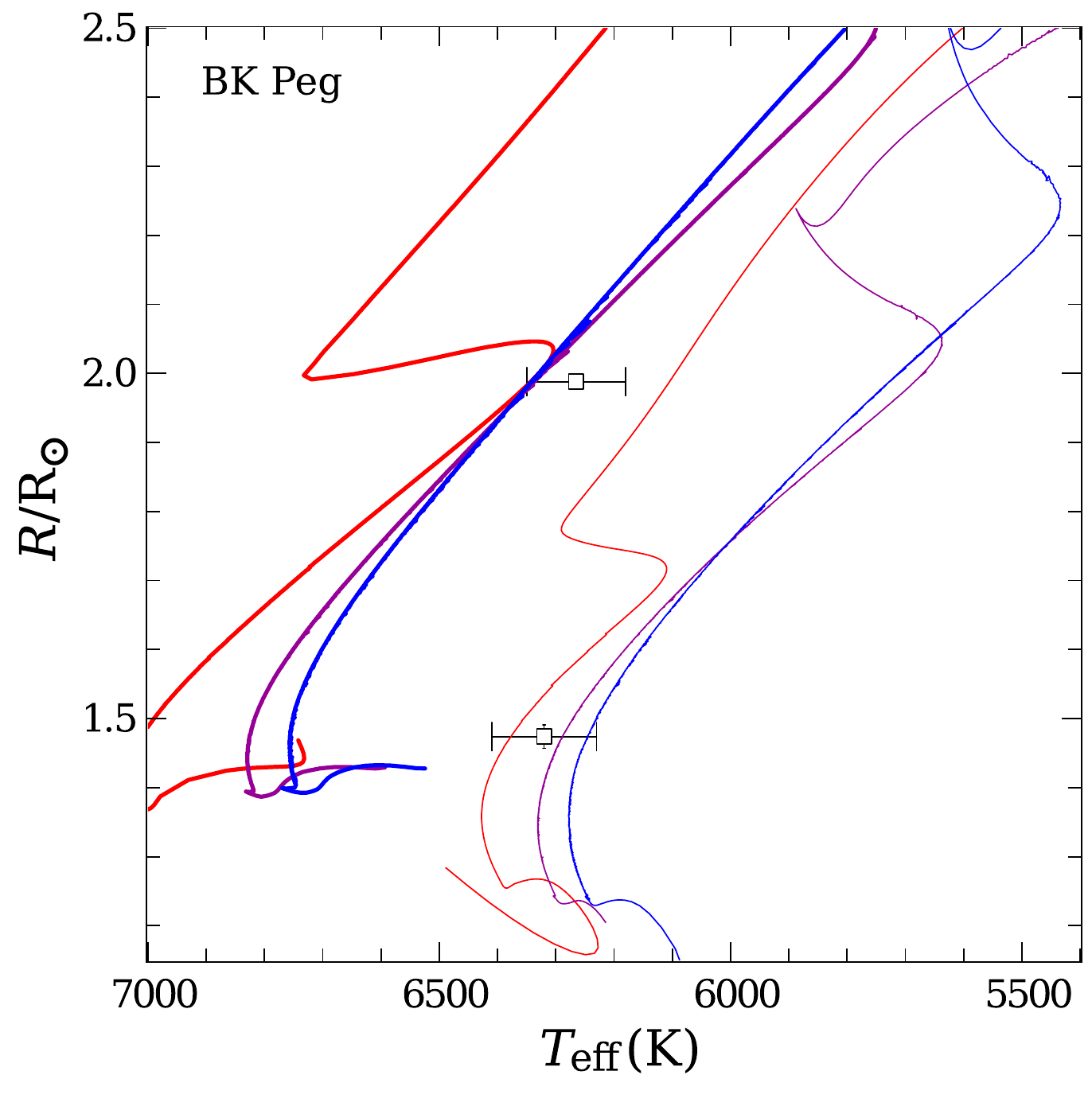}
  \caption{Evolution tracks of stellar models of BK~Peg.  Thick lines denote the primary models.  Error bars indicate the uncertainty in $R$ and $T_\text{eff}$ reported by \citet{2010A&A...516A..42C}.  The models have $\text{[Fe/H]} = -0.06$, $0.00$, and $0.01$; $f_\text{os} = 0.000$, $0.025$, and $0.040$; and $\alpha_\text{MLT} = 1.23$, $1.35$, and $1.35$; in red, magenta, and blue, respectively.}
  \label{figure_BKPeg}
\end{figure}

Three solutions for the system are presented in Figure~\ref{figure_BKPeg}.  Although there is a non-negligible mass difference between the two components (and therefore the potential for the `correct' amount of overshooting {for the two members to differ if there is a steep dependence of $f_\text{os}$ on stellar mass)} we again restricted our search to pairs with the same overshooting parameter.  Despite this, we were able to find solutions with a wide range of overshooting: $0.000 \leq f_\text{os} \leq 0.040$.  The models have a narrow metallicity range: $-0.06 \leq \text{[Fe/H]} \leq 0.01$, which although slightly higher than the spectroscopic value, is lower than the best fit from \citet{2018ApJ...859..100C}.

\begin{figure}
\includegraphics[width=\linewidth]{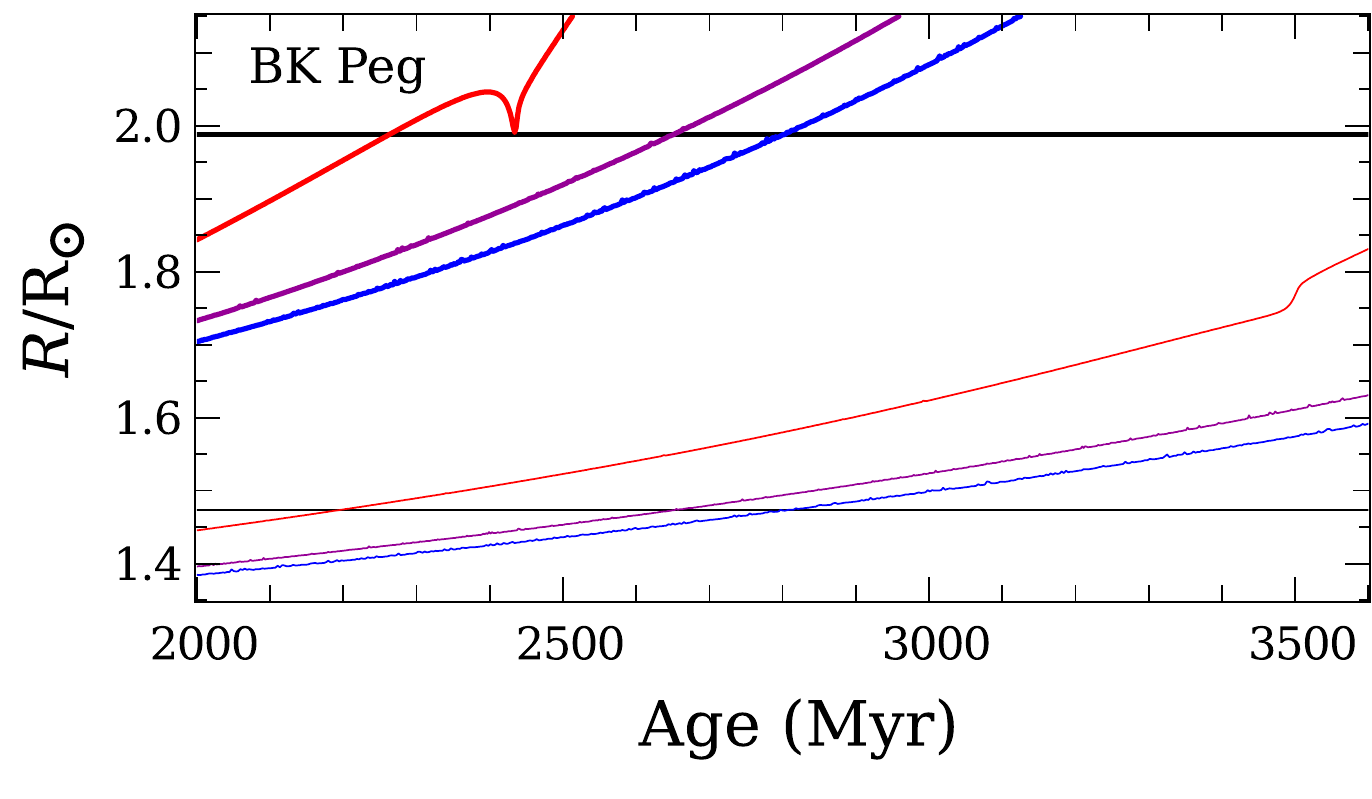}
  \caption{Evolution of radius of the BK~Peg models shown in Figure~\ref{figure_BKPeg}.  The colours are the same as Figure~\ref{figure_BKPeg} and the thick and thin horizontal lines show the primary and secondary radius, respectively.}
  \label{figure_BKPeg_R}
\end{figure}

In each of the solutions presented, we reduced the MLT mixing length compared with the initial default used in this paper $\alpha_\text{MLT} = 1.6$.  We still chose the same $\alpha_\text{MLT}$ for both members of each pair.  This reduction speeds the evolution of the radius of the secondary significantly more than for the primary and therefore ensures the two components are coeval.  This is possible because $\alpha_\text{MLT}$ has a contrasting influence on the evolution of the two components: the effect on age and $T_\text{eff}$ at a given radius is about four times larger for the secondary.  The good agreement between the ages of the two components of each pair is shown in Figure~\ref{figure_BKPeg_R}.  The reduction in $\alpha_\text{MLT}$ in this case may be justified because the convective envelopes are very thin: they encompass only $6\times10^{-5}\,\text{M}_\odot$ and $2 \times10^{-5}\,\text{M}_\odot$ in the primary and secondary models, respectively.

When the amount of overshooting increases, the required metallicity and $\alpha_\text{MLT}$ increase and decrease, respectively.  In addition to the metallicity increasing further above the spectroscopic value, as $f_\text{os}$ increases the $T_\text{eff}$ of the best fit primary becomes hotter and the secondary cooler, suggesting $f_\text{os} = 0.040$ is reasonably close to the upper limit.  Like the previous two low mass systems, AY Cam and HD 187669, we were not able to establish any meaningful constraints for $f_\text{os}$ using BK Peg.

\section{Models of helium burning stars}

\subsection{OGLE-LMC-ECL-26122}
\label{sec:OGLE-LMC-ECL-26122}

OGLE-LMC-ECL-26122 is an evolved system with 3.593$\,\text{M}_\odot$ and 3.411$\,\text{M}_\odot$ stars \citep{2013Natur.495...76P}.  \citet{2017ApJ...849...18C} found best fit models with $f_\text{os}=0.0190$ and $f_\text{os}=0.0170$ for the primary and secondary, respectively.  Importantly, they required two different MLT mixing length parameters, $\alpha_\text{MLT}=1.80$ and $\alpha_\text{MLT}=2.13$.

\begin{figure}
\includegraphics[width=\linewidth]{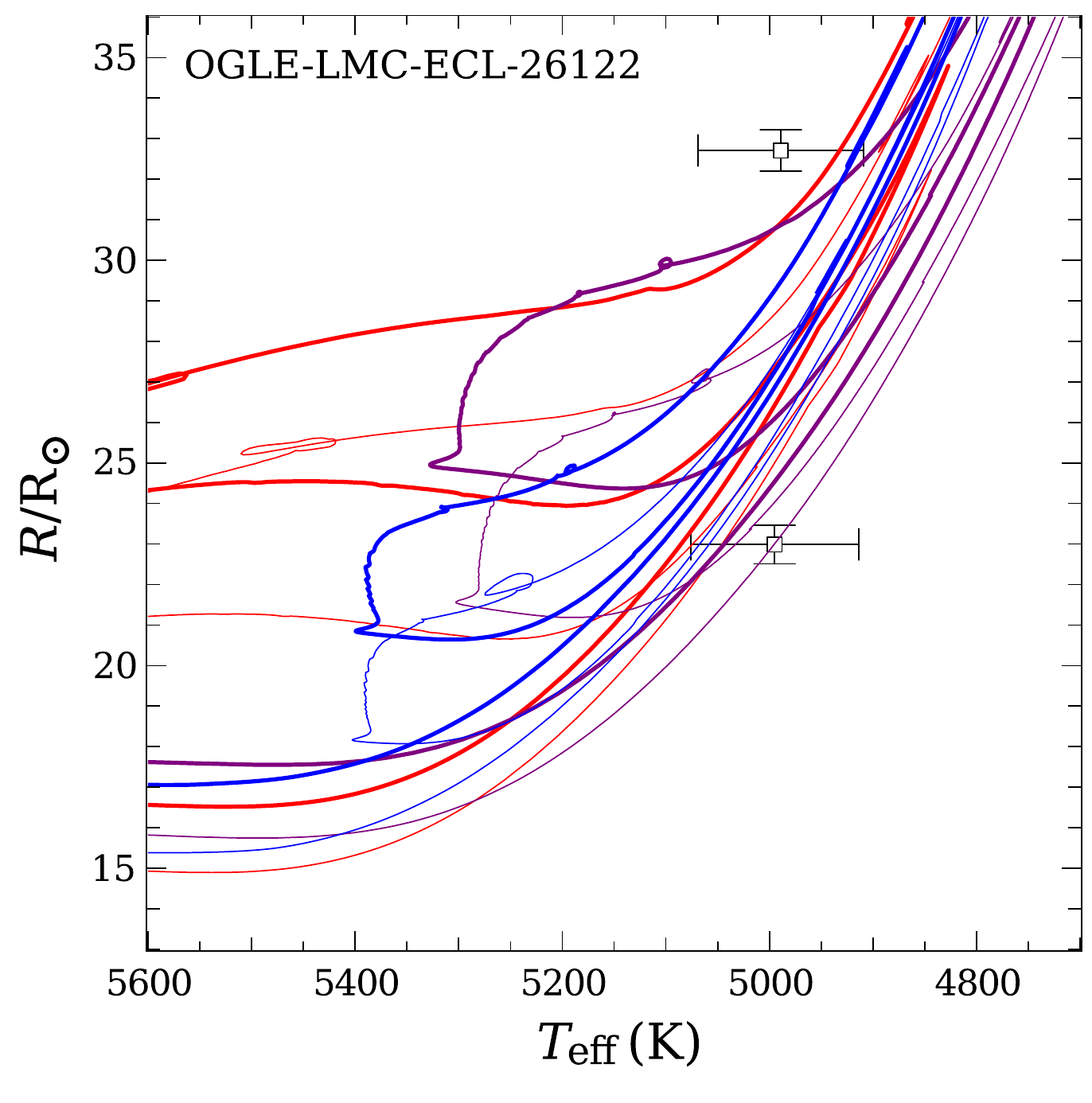}
  \caption{Evolution tracks of stellar models of OGLE-LMC-ECL-26122.  Thick lines denote the primary models.  Error bars indicate the uncertainty in $R$ and $T_\text{eff}$ reported by \citet{2013Natur.495...76P}.  The models have $0.005 \leq f_\text{os} \leq 0.020$ and $-0.50 \leq \text{[Fe/H]} \leq -0.15$ (red curves show models with the lowest $f_\text{os}$ and [Fe/H]). }
  \label{figure_OGLE-LMC-ECL-26122}
\end{figure}

Figure~\ref{figure_OGLE-LMC-ECL-26122} shows the evolution of three pairs of models, with $f_\text{os} = 0.005$, $f_\text{os} = 0.010$, and $f_\text{os} = 0.020$ with $\text{[Fe/H]} = -0.50$, $\text{[Fe/H]} = -0.30$ and $\text{[Fe/H]} = -0.15$, respectively.  Although two of these are below the spectroscopic determination of $\text{[Fe/H]} = -0.15 \pm 0.10$ by \citet{2013Natur.495...76P}, our lowest metallicity fit is only marginally more metal-poor than the best fit from \citet{2017ApJ...849...18C}, $Z = 0.0050$.

\begin{figure}
\includegraphics[width=\linewidth]{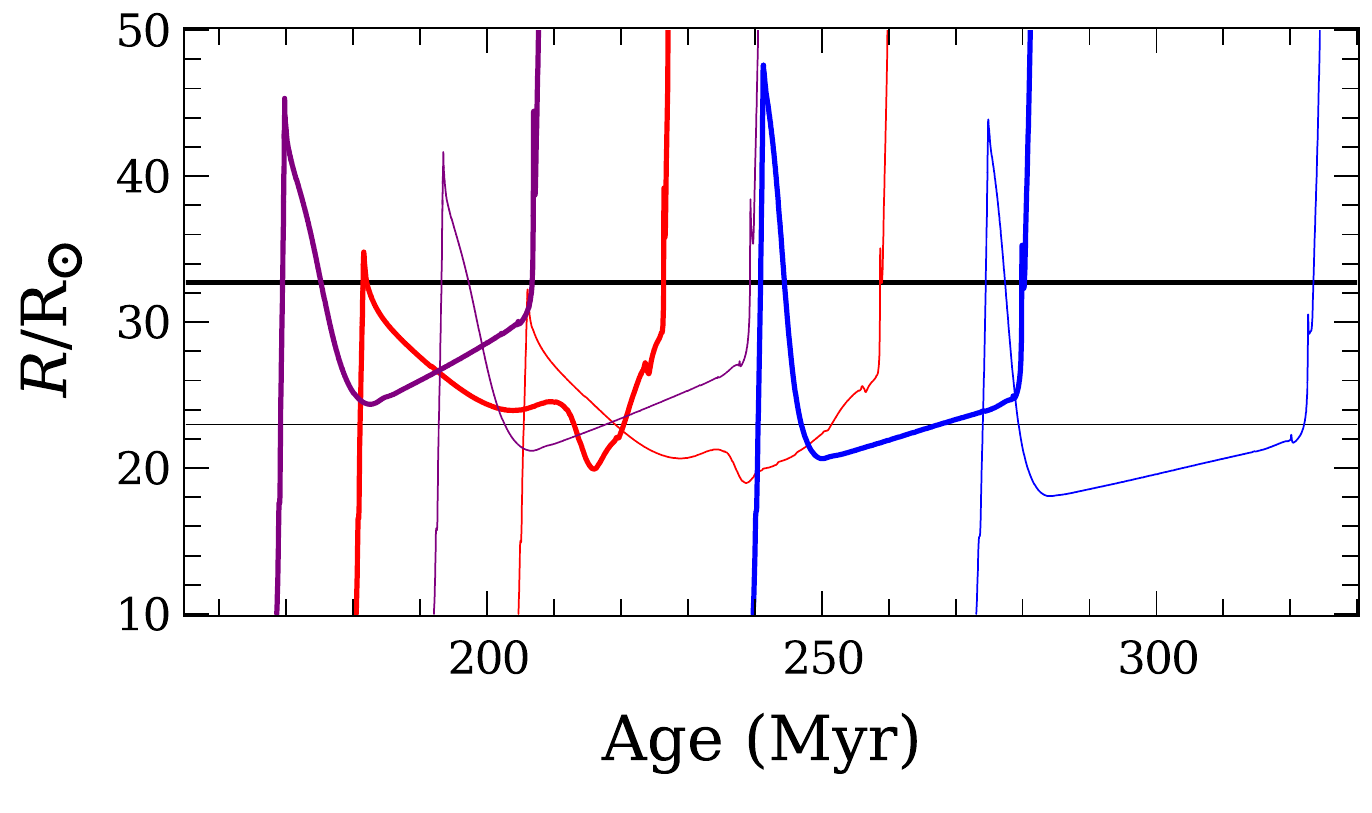}
  \caption{Evolution of radius of the OGLE-LMC-ECL-26122 models shown in Figure~\ref{figure_OGLE-LMC-ECL-26122}.  The colours are the same as Figure~\ref{figure_OGLE-LMC-ECL-26122} and the thick and thin horizontal lines show the primary and secondary radius, respectively.}
  \label{figure_OGLE-LMC-ECL-26122_R}
\end{figure}

The secondaries twice pass through the correct position in $R - T_\text{eff}$ space: once during the ascent of the RGB and secondly at the beginning of the core helium burning.  The primaries only once pass through the correct position in the $R - T_\text{eff}$ diagram: during the beginning of the ascent of the asymptotic giant branch, which is a relatively fast phase of evolution.  Figure~\ref{figure_OGLE-LMC-ECL-26122_R} shows that the age difference for each match is less than five per cent.

Each of our solutions has the two components in relatively rapid phases of evolution: the primary is an early-AGB star ascending the giant branch and the secondary is either an RGB or early-core helium burning star.  The respective phases for the two members, however, coincide in age over a large range of $f_\text{os}$.  The determination of the same $T_\text{eff}$ for both components presents a small challenge because the primary models tend to be systematically cooler than observed and the secondary models hotter than observed.  Precise metallicity constraints would help to further refine the permissible range of $f_\text{os}$ because there is a degeneracy between metallicity and overshooting: the luminosity increase from higher $f_\text{os}$ can be compensated with a reduction in luminosity from an increase in $\text{[Fe/H]}$.

\subsection{LMC-562.05-9009}
\label{sec:LMC-562.05-9009}

\begin{figure}
\includegraphics[width=\linewidth]{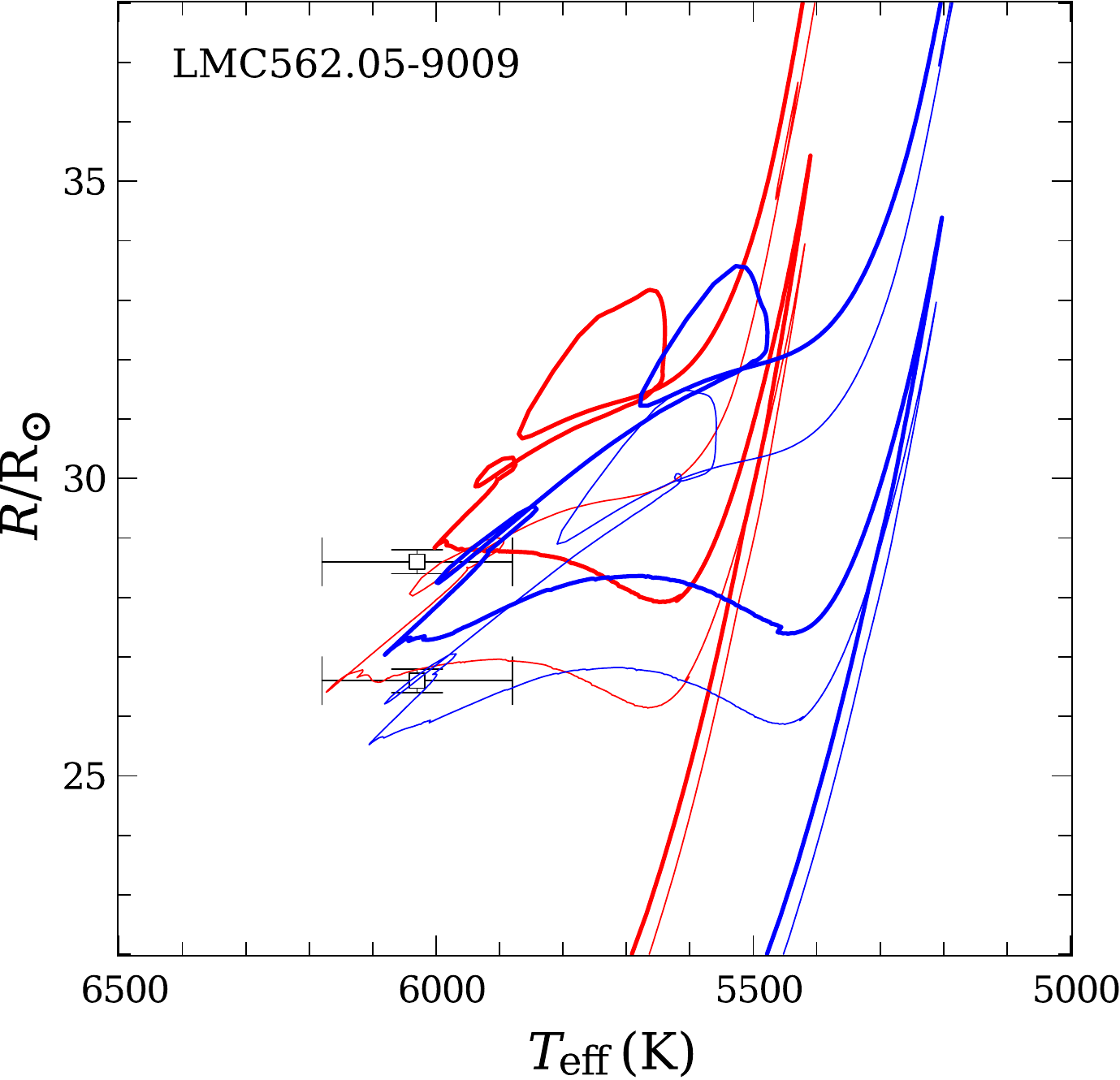}
  \caption{Evolution tracks of stellar models of LMC-562.05-9009.  Thick lines denote the primary models.  Error bars indicate the uncertainty in $R$ and $T_\text{eff}$ reported by \citet{2015ApJ...815...28G}.  The models have $f_\text{os} = 0.009$ (red) and $f_\text{os} = 0.014$ (blue) with $\text{[Fe/H]} = -0.70$.}
  \label{figure_LMC562.05.9009}
\end{figure}

LMC-562.05-9009 is a pair of core helium burning stars with a 3.7$\,\text{M}_\odot$ primary and a 3.6$\,\text{M}_\odot$ secondary \citep{2015ApJ...815...28G}.  \citet{2017ApJ...849...18C} found best fits of $f_\text{os}=0.0132$ and $f_\text{os} = 0.0128$, respectively for the two components, using the A09 mixture.  Two matching pairs we calculated, with $f_\text{os} = 0.007$ and $f_\text{os} = 0.018$ are shown in Figure~\ref{figure_LMC562.05.9009}.  The models have metallicity, $\text{[Fe/H]} = -0.7$, consistent with the $Z = 0.0025$ value for the best fit models from \citet{2017ApJ...849...18C}.  The range in $f_\text{os}$ is possible by increasing the MLT mixing length for models with more overshooting, specifically we used $\alpha_\text{MLT} = 2.4$ and $\alpha_\text{MLT} = 3.0$ for these two pairs of models.  The plausibility of these values for $\alpha_\text{MLT}$ may be explored by comparing predictions with observations of open clusters of a similar age to the LMC-562.05-9009 system.

\begin{figure}
\includegraphics[width=\linewidth]{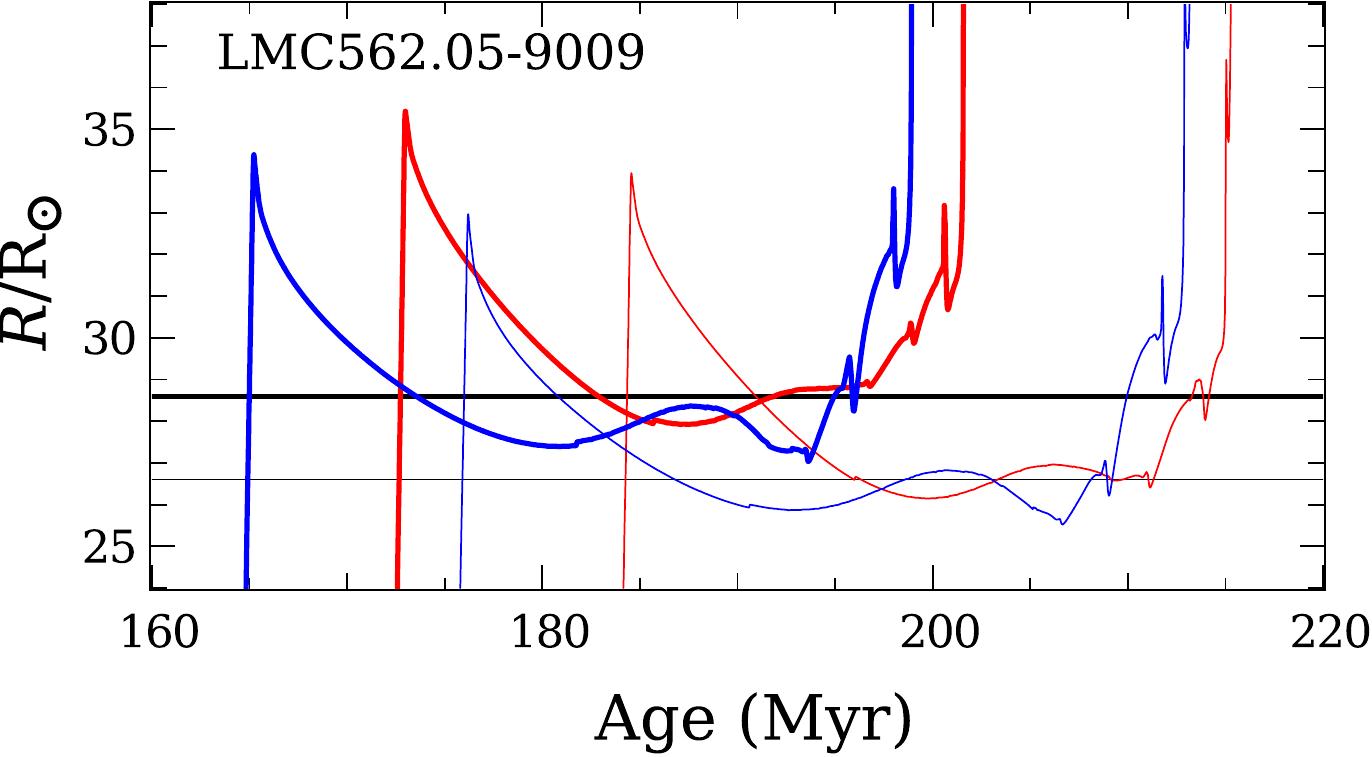}
  \caption{Evolution of radius of the LMC-562.05-9009 models shown in Figure~\ref{figure_LMC562.05.9009}.  The colours are the same as Figure~\ref{figure_LMC562.05.9009} and the thick and thin horizontal lines show the primary and secondary radius, respectively.}
  \label{figure_LMC562.05.9009_R}
\end{figure}

Figure~\ref{figure_LMC562.05.9009_R} shows that in both pairs the secondary model is older than the primary.  In both cases, the age difference between the two components is about 8\,Myr (it appears less in Figure~\ref{figure_LMC562.05.9009_R} because the first two times that the secondary attains the correct radius during core helium burning, $T_\text{eff}$ is still too cool). Uncertainties in the mass ratio, $0.974 \pm 0.004$, $0.973 \pm 0.005$, and $0.965 \pm 0.005$ for the three solutions from \citet{2015ApJ...815...28G}, could perhaps explain about 3\,Myr of this difference.  \citet{2015ApJ...815...28G} found that unlike the primary, the secondary does not pulsate, suggesting it is outside the red edge of the instability strip.  If the secondary is indeed cooler than the primary, the range of acceptable solutions to the system would be widened and the age difference between the two components reduced.

We stress that we have presented two possible solutions but have not completely explored the parameter space, which in this complicated case includes metallicity, overshooting, MLT mixing length, possible differences between $f_\text{os}$ and $\alpha_\text{MLT}$ for the two components, and uncertainties in the helium-burning reaction rates.

\subsection{CEP-0227}
\label{sec:CEP0227}

\begin{figure}
\includegraphics[width=\linewidth]{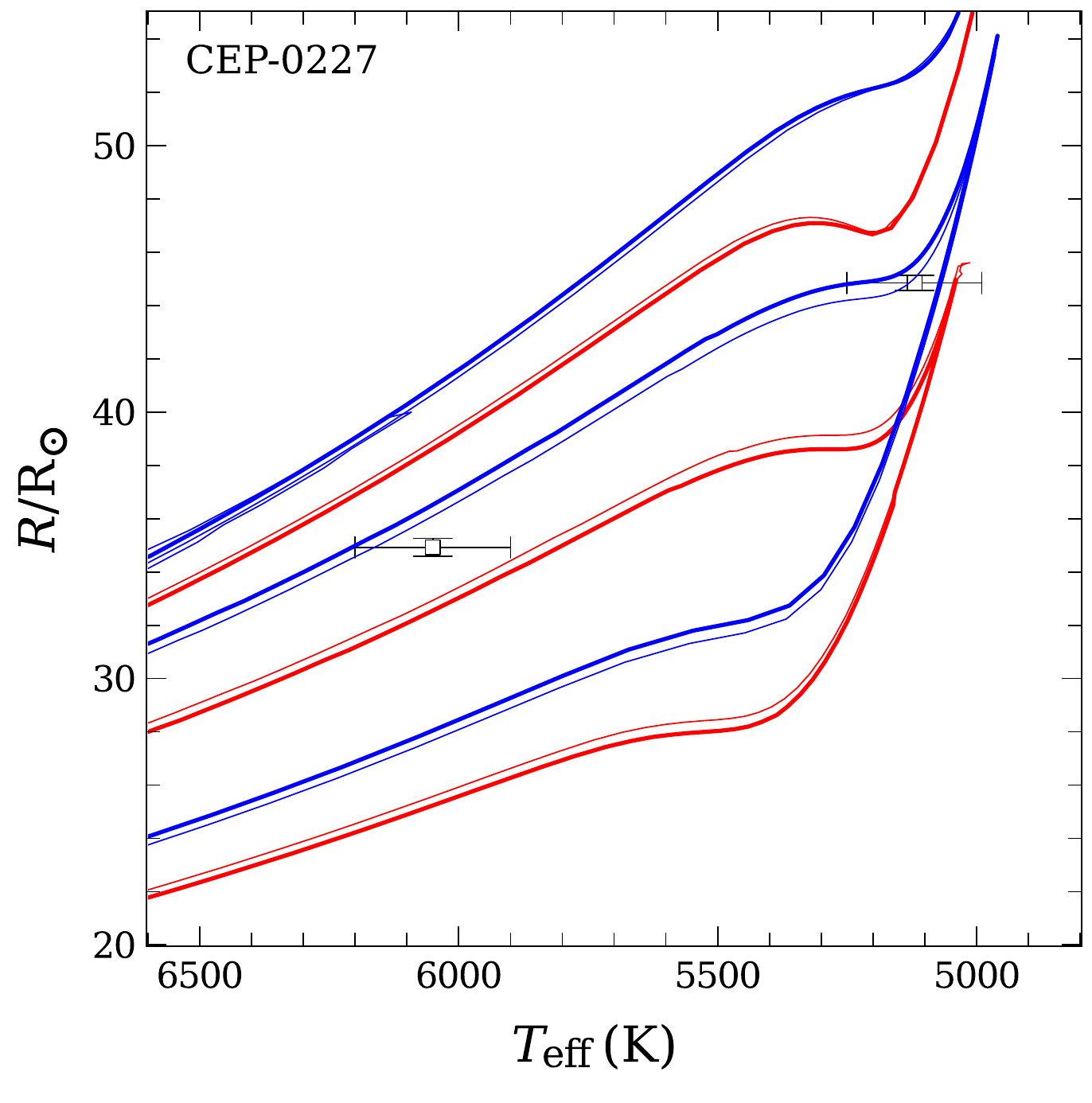}
  \caption{Evolution tracks of stellar models of CEP-0227.  The models have $f_\text{os} = 0.011$ (red) and $f_\text{os} = 0.018$, and the same metallicity $\text{[Fe/H]} = -1.0$.  Thick lines denote the primary models.  Error bars indicate the uncertainty in $R$ and $T_\text{eff}$ reported by \citet{2013MNRAS.436..953P}.}
  \label{figure_CEP-0227}
\end{figure}

CEP-0227 is a well-studied system with a $4.165\,\text{M}_\odot$ primary and a $4.134\,\text{M}_\odot$ secondary \citep{2013MNRAS.436..953P}.  \citet{2017ApJ...849...18C} found best fit models with $f_\text{os}=0.0150$ and $Z = 0.0022$ for both components, using the A09 mixture.  \citet{2017A&A...608A..62H} needed overshooting (and used their standard parameter $f_\text{os} = 0.02$) to fit the system.  In Figure~\ref{figure_CEP-0227} we present two sets of models spanning a wide range of overshooting: $f_\text{os} = 0.011$ and $f_\text{os} = 0.018$, both with metallicity $\text{[Fe/H]} = -1.0$.  The metallicity agrees reasonably well with the best fit by \citet{2017ApJ...849...18C}, but we note that in this case the metallicity is not very important: an increases of $\Delta  \text{[Fe/H]} = +0.20$ decreases $T_\text{eff}$ for the secondary by only 60\,K and increases the primary $T_\text{eff}$ by less than 10\,K.  Additionally, adjusting the MLT mixing length has no effect on the fit for the secondary because there is no convective envelope.  {More solutions are possible by altering only the initial helium abundance (from $Y = 0.25$): we have verified that exploring models with $0.245 \leq Y \leq 0.28$, for example, expands the range of valid overshooting parameters to $0.009 \leq f_\text{os} \leq 0.019$.}

\begin{figure}
\includegraphics[width=\linewidth]{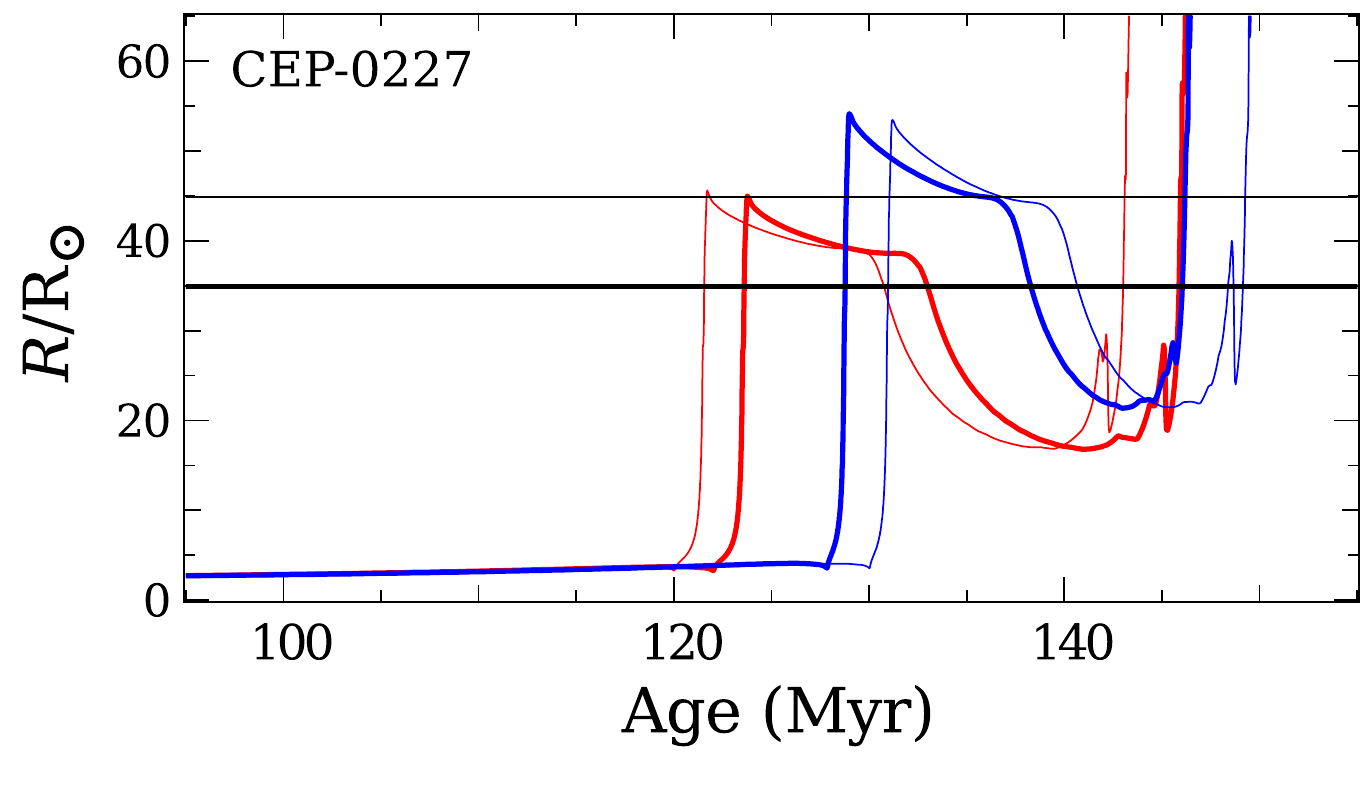}
  \caption{Evolution of radius of the CEP-0227 models shown in Figure~\ref{figure_CEP-0227}.  The colours are the same as Figure~\ref{figure_CEP-0227} and the thick and thin horizontal lines show the primary and secondary radius, respectively.}
  \label{figure_CEP-0227_R}
\end{figure}

Figure~\ref{figure_CEP-0227_R} shows that both the $f_\text{os} = 0.011$ and $f_\text{os} = 0.018$ evolution tracks pass through the observed data points during relatively slow phases of evolution.  The secondary for the $f_\text{os} = 0.011$ pair passes through the required $T_\text{eff}$ and $R$ at the tip of the RGB/very beginning of core helium burning whereas the $f_\text{os} = 0.018$ secondary does this after about 5\,Myr of core helium burning.  We note that it is possible to find a wider range of solutions if we allow a larger age discrepancy: there is a solution with both components in the latter part of core helium burning, and the secondary appearing to be in a more advanced stage of evolution.  Although this seems unlikely, it is a possible result from the (model-dependent) stochastic mixing episodes {known as core breathing pulses} that can occur late in core helium burning, which we discuss in Section~\ref{sec:cheb}.  These solutions, however, also have the disadvantage that radius evolution is faster in the later part of core helium burning, making the observation of such stars less probable.


\subsection{A note on the more evolved stars in the sample}
\label{sec:cheb}

We have find much tighter constraints for the evolved systems than for the main sequence systems in this paper.  This difference may result in part from neglecting uncertainties in the post-main sequence phases such as core helium burning overshooting and reaction rates, red giant branch mass loss, the MLT mixing length, and the possibility of multiple passes through the same $R$ and $T_\text{eff}$.  Our manual search, with a limited exploration of the parameter space of the models, may also limit the range of overshooting parameters found to be compatible with the observations.  We mention some of the ways this may influence our findings below.

Observational evidence from cluster star counts \citep[e.g.][]{1983A&A...128...94B,1986MmSAI..57..411B,1989ApJ...340..241C, 2016MNRAS.456.3866C} and asteroseismology \citep[e.g.][]{2013ApJ...766..118M,2015MNRAS.453.2290B,2015MNRAS.452..123C,2017MNRAS.469.4718B,2017MNRAS.472.4900C} unambiguously implies there is a need for overshooting during the core helium burning phase.  During the bulk of this phase, any non-negligible amount of overshoot initiates a feedback process that ensures that the evolution is relatively insensitive to the precise amount of overshooting or the particular scheme employed.  Later on, when core breathing pulses become important, the numerical treatment of mixing can strongly influence the evolution.  During the entire phase, the type of scheme or its implementation may still control whether a particular evolution sequence satisfies tight constraints on $R$ and $T_\text{eff}$.  Overshooting can extend the blue loops in the HR diagram and increase the luminosity towards the end of core helium burning.  These effects have been explored using the {\sc monstar} code in great detail \citep{2015MNRAS.452..123C,2016MNRAS.456.3866C,2017MNRAS.472.4900C}.

Core helium burning overshooting appears to be strongly favoured in the case of OGLE-LMC-ECL-26122.  If $f_\text{os}$ is large enough, say around $f_\text{os} = 0.02$, overshooting is required to lengthen the core helium burning phase so that the two components undergo it at the same time.  This is not strictly necessary, however, because the secondary passes through the correct $R$ and $T_\text{eff}$ during the ascent of the red giant branch, but this is a very rapid phase: the radius evolves through the 1$\sigma$ uncertainty $22.99 \pm 0.48\,\text{R}_\odot$ in less than 50\,kyr, suggesting it is an unlikely solution.

\begin{figure}
\includegraphics[width=\linewidth]{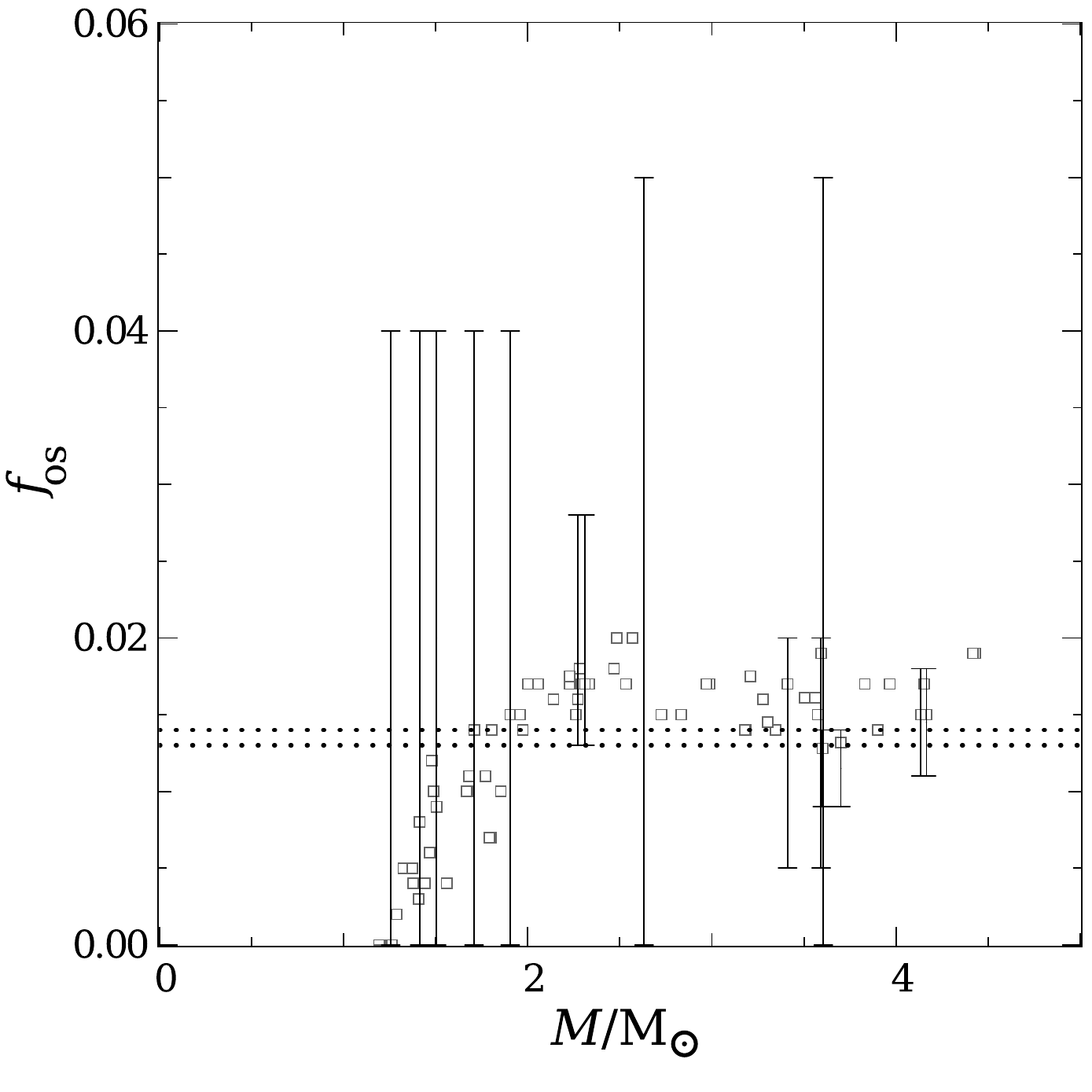}
  \caption{Comparison between the trend of overshoot with stellar mass for models in this paper and that reported by \citet{2017ApJ...849...18C} and \citet{2018ApJ...859..100C}.  The dotted lines show the range of overshoot found to be acceptable solutions for all systems in this study (and where members of each pair have the same overshooting parameter and MLT mixing length parameters.  The square symbols show the best fits from \citet{2017ApJ...849...18C} using the A09 mixture.}
  \label{figure_results_summary}
\end{figure}

\setlength{\tabcolsep}{3pt}
\begin{table*}

\begin{center}
  \caption{{Observational constraints and ranges of acceptable model parameters.  References for the observations are given in the text.}}
  \label{table_results_summary}
 \footnotesize
\begin{tabular}{lr@{}lr@{}lr@{}lrrrrrrrrr}
\toprule
   & \multicolumn{7}{c}{ Observervational constraints } & & \multicolumn{7}{c}{ Model parameters \vspace{0.07cm} }  \\ \cline{2-8} \cline{10-16}
   \multicolumn{15}{c}{\vspace{-0.15cm}}\\
  \multicolumn{8}{c}{ }  & & \multicolumn{2}{c}{$f_\text{os}$ \vspace{0.1cm}} & \multicolumn{2}{c}{$\text{[Fe/H]}$} &  \multicolumn{2}{c}{$\alpha_\text{MLT}$} & \\

Name  & \multicolumn{2}{c}{$M/\text{M}_\odot$} & \multicolumn{2}{c}{$R/\text{R}_\odot$} & \multicolumn{2}{c}{$T_\text{eff}$ (K)} & \multicolumn{1}{c}{[Fe/H]} & & lower & upper & lower & upper & lower & upper & \multicolumn{1}{c}{$Y$} \\
\midrule
SZ~Cen              & $2.311 \pm$&$ 0.026$ & $4.556 \pm$&$ 0.032$ & $8100 \pm$&$ 300$  &        -         & & 0.013 & 0.028 & -0.20 & -0.25 & 1.60 & 1.60 & 0.25 \\
$\dots$             & $2.272 \pm$&$ 0.021$ & $3.626 \pm$&$ 0.026$ & $8380 \pm$&$ 300$  &        -         & & 0.013 & 0.028 & -0.20 & -0.25 & 1.60 & 1.60 & 0.25 \\
AY~Cam              & $1.905 \pm$&$ 0.040$ & $2.772 \pm$&$ 0.020$ & $7250 \pm$&$ 100$  &        -         & & 0.000 & 0.040 &  0.00 &  0.10 & 1.60 & 1.60 & 0.26 \\
$\dots$             & $1.709 \pm$&$ 0.036$ & $2.026 \pm$&$ 0.017$ & $7395 \pm$&$ 100$  &        -         & & 0.000 & 0.040 &  0.00 &  0.10 & 1.60 & 1.60 & 0.26 \\
OGLE-LMC-ECL-26122  & $3.593 \pm$&$ 0.055$ & $32.71 \pm$&$ 0.51$  & $4989 \pm$&$ 80$   & $-0.15 \pm 0.10$ & & 0.005 & 0.020 & -0.50 & -0.15 & 1.90 & 2.20 & 0.25 \\ 
$\dots$             & $3.411 \pm$&$ 0.047$ & $22.99 \pm$&$ 0.48$  & $4995 \pm$&$ 81$   & $\dots$          & & 0.005 & 0.020 & -0.50 & -0.15 & 1.90 & 2.20 & 0.25 \\ 
LMC-562.05-9009     & $3.700 \pm$&$ 0.03$  & $28.6 \pm$&$ 0.2$    & $6030 \pm$&$ 150$  &        -         & & 0.009 & 0.014 & -0.70 & -0.70 & 2.40 & 3.00 & 0.25 \\
$\dots$             & $3.600 \pm$&$ 0.03$  & $26.6 \pm$&$ 0.2$    & $6030 \pm$&$ 150$  &        -         & & 0.009 & 0.014 & -0.70 & -0.70 & 2.40 & 3.00 & 0.25 \\
HD~187669           & $1.505 \pm$&$ 0.004$ & $22.62 \pm$&$ 0.50$  & $4330 \pm$&$ 70$   & $-0.25 \pm 0.10$ & & 0.000 & 0.040 & -0.25 & -0.25 & 1.60 & 1.60 & 0.26 \\
$\dots$             & $1.504 \pm$&$ 0.004$ & $11.33 \pm$&$ 0.28$  & $4650 \pm$&$ 80$   & $\dots$          & & 0.000 & 0.040 & -0.25 & -0.25 & 1.60 & 1.60 & 0.26 \\ 
CEP-0227            & $4.165 \pm$&$ 0.032$ & $34.92 \pm$&$ 0.34$  & $6050 \pm$&$ 160$  &        -         & & 0.011 & 0.018 & -1.00 & -1.00 & 2.00 & 2.00 & 0.25 \\
$\dots$             & $4.134 \pm$&$ 0.037$ & $44.85 \pm$&$ 0.29$  & $5120 \pm$&$ 130$  &        -         & & 0.011 & 0.018 & -1.00 & -1.00 & 2.00 & 2.00 & 0.25 \\
$\chi^2$ Hya        & $3.605 \pm$&$ 0.078$ & $4.390 \pm$&$ 0.039$ & $11750 \pm$&$ 190$ &        -         & & 0.000 & 0.050 & -0.15 &  0.00 & 1.60 & 1.60 & 0.26 \\ 
$\dots$             & $2.632 \pm$&$ 0.049$ & $2.159 \pm$&$ 0.030$ & $11100 \pm$&$ 230$ &        -         & & 0.000 & 0.050 & -0.15 &  0.00 & 1.60 & 1.60 & 0.26 \\
BK Peg              & $1.414 \pm$&$ 0.007$ & $1.988 \pm$&$ 0.008$ & $6265 \pm$&$ 85$   & $-0.12 \pm 0.07$ & & 0.000 & 0.040 & -0.06 &  0.05 & 1.23 & 1.35 & 0.26 \\
$\dots$             & $1.257 \pm$&$ 0.005$ & $1.474 \pm$&$ 0.017$ & $6320 \pm$&$ 90$   & $\dots$          & & 0.000 & 0.040 & -0.06 &  0.05 & 1.23 & 1.35 & 0.26 \\
\bottomrule
\end{tabular}
\normalsize
\end{center}
\normalsize
\end{table*}

\section{Summary and conclusions}

In response to recent findings that the amount of main sequence overshooting required to explain the observations of double-lined eclipsing binary stars is strongly dependent on stellar mass, we have conducted a detailed exploration of a sample of such systems and tested the sensitivity of the results to some important uncertainties.  We took a representative selection of eight eclipsing binary systems, covering a wide mass range and including stars in various phases of evolution, from the samples used by \citet{2016A&A...592A..15C,2017ApJ...849...18C,2018ApJ...859..100C}.  We modelled overshooting (and any other mechanisms for mixing near the boundary of the convective core) by varying the free parameter $f_\text{os}$ in the scheme from \citet{1997A&A...324L..81H}, where there is an exponential decay in the diffusion coefficient in formally stable regions.

We investigated an array of models for each system to establish a range of overshooting parameters that yielded acceptable solutions according to effective temperature, radius, age, and metallicity constraints.  These results are presented in Table~\ref{table_results_summary}.  In general, our results are indicative of the range of overshooting consistent with the observations but do not necessarily reach the possible extremes.  We compare our determinations for the amount of overshooting with \citet{2017ApJ...849...18C,2018ApJ...859..100C} in Figure~\ref{figure_results_summary}.  Our findings are usually consistent with the best fit models from \citet{2017ApJ...849...18C,2018ApJ...859..100C} but we find a large range of acceptable $f_\text{os}$ that makes it very difficult to detect any trend with mass.  We confirm earlier results that the evidence strongly supports the requirement for overshooting in models of stars with $M \gtrsim 2\,\text{M}_\odot$.  We could match all of the eight pairs with $0.013 \leq f_\text{os} \leq 0.014 $ (and seven of the pairs with $0.013 \leq f_\text{os} \leq 0.018 $), which is remarkably consistent with the range of best fit $f_\text{os} \approx 0.016$ found by \citet{2017ApJ...849...18C} for stars with $M > 2\,\text{M}_\odot$.

None of the five DLEB pairs of main sequence or subgiant stars were particularly useful for constraining core overshooting.  We were, however, able to more tightly constrain the overshooting parameter in models in later phases of evolution.  Unfortunately, this presents new challenges because the radius and effective temperature evolution of those models are more strongly dependent on the mixing length parameter and metallicity, and stars can pass through the same place in the HR diagram multiple times, which complicates the search for the most favourable parameters.  We have shown that in most cases a valid solution exists with a range of overshooting parameter, even without conducting an exhaustive search of the parameter space, which additionally includes metallicity, helium abundance, possible discrepancies between $f_\text{os}$ or $\alpha_\text{MLT}$ for the two components, and uncertainties in the helium-burning reaction rates for evolved systems.  We also caution that in this study we have not formally weighted the solution likelihoods where it may be possible, by considering the duration of the windows of valid solutions with each combination of parameters, for example.

In their recent paper, \citet{2018arXiv180307058V} raised the question of whether their conclusions about the difficulty of precisely constraining the overshooting from an eclipsing binary pair apply generally.  We have identified that in most cases it is indeed difficult to definitively determine the extent of overshooting from the available measurements of stellar masses, radii, and effective temperatures.  In many examples, the allowed range of the overshooting parameters could be reduced with more precise determinations of effective temperature and metallicity.  The situation may also be helped by complementary approaches such as asteroseismology and hydrodynamical models which are now being applied to the same problem.  We also wish to emphasize the value of the recent approach of \citet{2016A&A...592A..15C} where models for large numbers of systems are assessed together, especially as observations improve in both quantity and quality, which will reduce the uncertainties in each specific case and therefore overall.

Overall, we do not find evidence to support a mass dependence for the amount of overshooting, other than that it is necessary for models with mass above about 2\,$\text{M}_\odot$.  We find that a constant overshooting parameter provides an adequate fit to the data.

\section*{Acknowledgements}

This project was supported by the European Research Council through ERC AdG No. 320478-TOFU and the STFC Consolidated Grant ST/R000395/1.

\footnotesize{
 \bibliographystyle{aa}
  \bibliography{eclipsing_binary_paper}

\begin{thebibliography}{54}
\expandafter\ifx\csname natexlab\endcsname\relax\def\natexlab#1{#1}\fi

\bibitem[{{Andersen}(1975)}]{1975A&A....45..203A}
{Andersen}, J. 1975, \aap, 45, 203

\bibitem[{{Andersen}(1991)}]{1991A&ARv...3...91A}
{Andersen}, J. 1991, \aapr, 3, 91

\bibitem[{{Angulo} {et~al.}(1999){Angulo}, {Arnould}, {Rayet}, {Descouvemont},
  {Baye}, {Leclercq-Willain}, {Coc}, {Barhoumi}, {Aguer}, {Rolfs}, {Kunz},
  {Hammer}, {Mayer}, {Paradellis}, {Kossionides}, {Chronidou}, {Spyrou},
  {degl'Innocenti}, {Fiorentini}, {Ricci}, {Zavatarelli}, {Providencia},
  {Wolters}, {Soares}, {Grama}, {Rahighi}, {Shotter}, \& {Lamehi
  Rachti}}]{1999NuPhA.656....3A}
{Angulo}, C., {Arnould}, M., {Rayet}, M., {et~al.} 1999, Nuclear Physics A,
  656, 3

\bibitem[{{Asplund} {et~al.}(2009){Asplund}, {Grevesse}, {Sauval}, \&
  {Scott}}]{2009ARA&A..47..481A}
{Asplund}, M., {Grevesse}, N., {Sauval}, A.~J., \& {Scott}, P. 2009, \araa, 47,
  481

\bibitem[{{Bastian} \& {de Mink}(2009)}]{2009MNRAS.398L..11B}
{Bastian}, N. \& {de Mink}, S.~E. 2009, \mnras, 398, L11

\bibitem[{{Bossini} {et~al.}(2015){Bossini}, {Miglio}, {Salaris},
  {Pietrinferni}, {Montalb{\'a}n}, {Bressan}, {Noels}, {Cassisi}, {Girardi}, \&
  {Marigo}}]{2015MNRAS.453.2290B}
{Bossini}, D., {Miglio}, A., {Salaris}, M., {et~al.} 2015, \mnras, 453, 2290

\bibitem[{{Bossini} {et~al.}(2017){Bossini}, {Miglio}, {Salaris}, {Vrard},
  {Cassisi}, {Mosser}, {Montalb{\'a}n}, {Girardi}, {Noels}, {Bressan},
  {Pietrinferni}, \& {Tayar}}]{2017MNRAS.469.4718B}
{Bossini}, D., {Miglio}, A., {Salaris}, M., {et~al.} 2017, \mnras, 469, 4718

\bibitem[{{Bressan} {et~al.}(1986){Bressan}, {Bertelli}, \&
  {Chiosi}}]{1986MmSAI..57..411B}
{Bressan}, A., {Bertelli}, G., \& {Chiosi}, C. 1986, \memsai, 57, 411

\bibitem[{{Buzzoni} {et~al.}(1983){Buzzoni}, {Pecci}, {Buonanno}, \&
  {Corsi}}]{1983A&A...128...94B}
{Buzzoni}, A., {Pecci}, F.~F., {Buonanno}, R., \& {Corsi}, C.~E. 1983, \aap,
  128, 94

\bibitem[{{Campbell} \& {Lattanzio}(2008)}]{2008A&A...490..769C}
{Campbell}, S.~W. \& {Lattanzio}, J.~C. 2008, \aap, 490, 769

\bibitem[{{Caputo} {et~al.}(1989){Caputo}, {Chieffi}, {Tornambe}, {Castellani},
  \& {Pulone}}]{1989ApJ...340..241C}
{Caputo}, F., {Chieffi}, A., {Tornambe}, A., {Castellani}, V., \& {Pulone}, L.
  1989, \apj, 340, 241

\bibitem[{{Claret}(2007)}]{2007A&A...475.1019C}
{Claret}, A. 2007, \aap, 475, 1019

\bibitem[{{Claret} \& {Torres}(2016)}]{2016A&A...592A..15C}
{Claret}, A. \& {Torres}, G. 2016, \aap, 592, A15

\bibitem[{{Claret} \& {Torres}(2017)}]{2017ApJ...849...18C}
{Claret}, A. \& {Torres}, G. 2017, \apj, 849, 18

\bibitem[{{Claret} \& {Torres}(2018)}]{2018ApJ...859..100C}
{Claret}, A. \& {Torres}, G. 2018, \apj, 859, 100

\bibitem[{{Clausen} {et~al.}(2010){Clausen}, {Frandsen}, {Bruntt}, {Olsen},
  {Helt}, {Gregersen}, {Juncher}, \& {Krogstrup}}]{2010A&A...516A..42C}
{Clausen}, J.~V., {Frandsen}, S., {Bruntt}, H., {et~al.} 2010, \aap, 516, A42

\bibitem[{{Cogan}(1975)}]{1975ApJ...201..637C}
{Cogan}, B.~C. 1975, \apj, 201, 637

\bibitem[{{Constantino} {et~al.}(2014){Constantino}, {Campbell}, {Gil-Pons}, \&
  {Lattanzio}}]{2014ApJ...784...56C}
{Constantino}, T., {Campbell}, S., {Gil-Pons}, P., \& {Lattanzio}, J. 2014,
  \apj, 784, 56

\bibitem[{{Constantino} {et~al.}(2015){Constantino}, {Campbell},
  {Christensen-Dalsgaard}, {Lattanzio}, \& {Stello}}]{2015MNRAS.452..123C}
{Constantino}, T., {Campbell}, S.~W., {Christensen-Dalsgaard}, J., {Lattanzio},
  J.~C., \& {Stello}, D. 2015, \mnras, 452, 123

\bibitem[{{Constantino} {et~al.}(2017){Constantino}, {Campbell}, \&
  {Lattanzio}}]{2017MNRAS.472.4900C}
{Constantino}, T., {Campbell}, S.~W., \& {Lattanzio}, J.~C. 2017, \mnras, 472,
  4900

\bibitem[{{Constantino} {et~al.}(2016){Constantino}, {Campbell}, {Lattanzio},
  \& {van Duijneveldt}}]{2016MNRAS.456.3866C}
{Constantino}, T., {Campbell}, S.~W., {Lattanzio}, J.~C., \& {van Duijneveldt},
  A. 2016, \mnras, 456, 3866

\bibitem[{{Espinosa Lara} \& {Rieutord}(2011)}]{2011A&A...533A..43E}
{Espinosa Lara}, F. \& {Rieutord}, M. 2011, \aap, 533, A43

\bibitem[{{Freytag} {et~al.}(1996){Freytag}, {Ludwig}, \&
  {Steffen}}]{1996A&A...313..497F}
{Freytag}, B., {Ludwig}, H.-G., \& {Steffen}, M. 1996, \aap, 313, 497

\bibitem[{{Gieren} {et~al.}(2015){Gieren}, {Pilecki}, {Pietrzy{\'n}ski},
  {Graczyk}, {Udalski}, {Soszy{\'n}ski}, {Thompson}, {Prada Moroni}, {Smolec},
  {Konorski}, {G{\'o}rski}, {Karczmarek}, {Suchomska}, {Taormina}, {Gallenne},
  {Storm}, {Bono}, {Catelan}, {Szyma{\'n}ski}, {Koz{\l}owski}, {Pietrukowicz},
  {Wyrzykowski}, {Poleski}, {Skowron}, {Minniti}, {Ulaczyk}, {Mr{\'o}z},
  {Pawlak}, \& {Nardetto}}]{2015ApJ...815...28G}
{Gieren}, W., {Pilecki}, B., {Pietrzy{\'n}ski}, G., {et~al.} 2015, \apj, 815,
  28

\bibitem[{{Grevesse} \& {Sauval}(1998)}]{1998SSRv...85..161G}
{Grevesse}, N. \& {Sauval}, A.~J. 1998, \ssr, 85, 161

\bibitem[{{Gronbech} {et~al.}(1977){Gronbech}, {Gyldenkerne}, \&
  {Jorgensen}}]{1977A&A....55..401G}
{Gronbech}, B., {Gyldenkerne}, K., \& {Jorgensen}, H.~E. 1977, \aap, 55, 401

\bibitem[{{He{\l}miniak} {et~al.}(2015){He{\l}miniak}, {Graczyk}, {Konacki},
  {Pilecki}, {Ratajczak}, {Pietrzy{\'n}ski}, {Sybilski}, {Villanova}, {Gieren},
  {Pojma{\'n}ski}, {Konorski}, {Suchomska}, {Reichart}, {Ivarsen}, {Haislip},
  \& {LaCluyze}}]{2015MNRAS.448.1945H}
{He{\l}miniak}, K.~G., {Graczyk}, D., {Konacki}, M., {et~al.} 2015, \mnras,
  448, 1945

\bibitem[{{Herwig} {et~al.}(1997){Herwig}, {Bloecker}, {Schoenberner}, \& {El
  Eid}}]{1997A&A...324L..81H}
{Herwig}, F., {Bloecker}, T., {Schoenberner}, D., \& {El Eid}, M. 1997, \aap,
  324, L81

\bibitem[{{Higl} \& {Weiss}(2017)}]{2017A&A...608A..62H}
{Higl}, J. \& {Weiss}, A. 2017, \aap, 608, A62

\bibitem[{{Iglesias} \& {Rogers}(1996)}]{1996ApJ...464..943I}
{Iglesias}, C.~A. \& {Rogers}, F.~J. 1996, \apj, 464, 943

\bibitem[{{Maeder} \& {Meynet}(1988)}]{1988A&AS...76..411M}
{Maeder}, A. \& {Meynet}, G. 1988, \aaps, 76, 411

\bibitem[{{Maeder} \& {Meynet}(1989)}]{1989A&A...210..155M}
{Maeder}, A. \& {Meynet}, G. 1989, \aap, 210, 155

\bibitem[{{Marigo} \& {Aringer}(2009)}]{2009A&A...508.1539M}
{Marigo}, P. \& {Aringer}, B. 2009, \aap, 508, 1539

\bibitem[{{Meng} \& {Zhang}(2014)}]{2014ApJ...787..127M}
{Meng}, Y. \& {Zhang}, Q.~S. 2014, \apj, 787, 127

\bibitem[{{Montalb{\'a}n} {et~al.}(2013){Montalb{\'a}n}, {Miglio}, {Noels},
  {Dupret}, {Scuflaire}, \& {Ventura}}]{2013ApJ...766..118M}
{Montalb{\'a}n}, J., {Miglio}, A., {Noels}, A., {et~al.} 2013, \apj, 766, 118

\bibitem[{{Pietrzy{\'n}ski} {et~al.}(2013){Pietrzy{\'n}ski}, {Graczyk},
  {Gieren}, {Thompson}, {Pilecki}, {Udalski}, {Soszy{\'n}ski}, {Koz{\l}owski},
  {Konorski}, {Suchomska}, {Bono}, {Moroni}, {Villanova}, {Nardetto},
  {Bresolin}, {Kudritzki}, {Storm}, {Gallenne}, {Smolec}, {Minniti}, {Kubiak},
  {Szyma{\'n}ski}, {Poleski}, {Wyrzykowski}, {Ulaczyk}, {Pietrukowicz},
  {G{\'o}rski}, \& {Karczmarek}}]{2013Natur.495...76P}
{Pietrzy{\'n}ski}, G., {Graczyk}, D., {Gieren}, W., {et~al.} 2013, \nat, 495,
  76

\bibitem[{{Pilecki} {et~al.}(2013){Pilecki}, {Graczyk}, {Pietrzy{\'n}ski},
  {Gieren}, {Thompson}, {Freedman}, {Scowcroft}, {Madore}, {Udalski},
  {Soszy{\'n}ski}, {Konorski}, {Smolec}, {Nardetto}, {Bono}, {Prada Moroni},
  {Storm}, \& {Gallenne}}]{2013MNRAS.436..953P}
{Pilecki}, B., {Graczyk}, D., {Pietrzy{\'n}ski}, G., {et~al.} 2013, \mnras,
  436, 953

\bibitem[{{Pols} {et~al.}(1997){Pols}, {Tout}, {Schroder}, {Eggleton}, \&
  {Manners}}]{1997MNRAS.289..869P}
{Pols}, O.~R., {Tout}, C.~A., {Schroder}, K.-P., {Eggleton}, P.~P., \&
  {Manners}, J. 1997, \mnras, 289, 869

\bibitem[{{Ribas} {et~al.}(2000){Ribas}, {Jordi}, \&
  {Gim{\'e}nez}}]{2000MNRAS.318L..55R}
{Ribas}, I., {Jordi}, C., \& {Gim{\'e}nez}, {\'A}. 2000, \mnras, 318, L55

\bibitem[{{Rogers} \& {Nayfonov}(2002)}]{2002ApJ...576.1064R}
{Rogers}, F.~J. \& {Nayfonov}, A. 2002, \apj, 576, 1064

\bibitem[{{Roxburgh}(1965)}]{1965MNRAS.130..223R}
{Roxburgh}, I.~W. 1965, \mnras, 130, 223

\bibitem[{{Roxburgh}(1978)}]{1978A&A....65..281R}
{Roxburgh}, I.~W. 1978, \aap, 65, 281

\bibitem[{{Roxburgh}(1989)}]{1989A&A...211..361R}
{Roxburgh}, I.~W. 1989, \aap, 211, 361

\bibitem[{Roxburgh(1999)}]{10.1007/978-94-011-4497-1_11}
Roxburgh, I.~W. 1999, in The Non-Sleeping Universe, ed. M.~T. V.~T. Lago \&
  A.~Blanchard (Dordrecht: Springer Netherlands), 43--50

\bibitem[{{Schaller} {et~al.}(1992){Schaller}, {Schaerer}, {Meynet}, \&
  {Maeder}}]{1992A&AS...96..269S}
{Schaller}, G., {Schaerer}, D., {Meynet}, G., \& {Maeder}, A. 1992, \aaps, 96,
  269

\bibitem[{{Schroder} {et~al.}(1997){Schroder}, {Pols}, \&
  {Eggleton}}]{1997MNRAS.285..696S}
{Schroder}, K.-P., {Pols}, O.~R., \& {Eggleton}, P.~P. 1997, \mnras, 285, 696

\bibitem[{{Shaviv} \& {Salpeter}(1973)}]{1973ApJ...184..191S}
{Shaviv}, G. \& {Salpeter}, E.~E. 1973, \apj, 184, 191

\bibitem[{{Stancliffe} {et~al.}(2015){Stancliffe}, {Fossati}, {Passy}, \&
  {Schneider}}]{2015A&A...575A.117S}
{Stancliffe}, R.~J., {Fossati}, L., {Passy}, J.-C., \& {Schneider}, F.~R.~N.
  2015, \aap, 575, A117

\bibitem[{{Timmes} \& {Swesty}(2000)}]{2000ApJS..126..501T}
{Timmes}, F.~X. \& {Swesty}, F.~D. 2000, \apjs, 126, 501

\bibitem[{{Torres} {et~al.}(2010){Torres}, {Andersen}, \&
  {Gim{\'e}nez}}]{2010A&ARv..18...67T}
{Torres}, G., {Andersen}, J., \& {Gim{\'e}nez}, A. 2010, \aapr, 18, 67

\bibitem[{{Valle} {et~al.}(2016){Valle}, {Dell'Omodarme}, {Prada Moroni}, \&
  {Degl'Innocenti}}]{2016A&A...587A..16V}
{Valle}, G., {Dell'Omodarme}, M., {Prada Moroni}, P.~G., \& {Degl'Innocenti},
  S. 2016, \aap, 587, A16

\bibitem[{{Valle} {et~al.}(2018){Valle}, {Dell'Omodarme}, {Prada Moroni}, \&
  {Degl'Innocenti}}]{2018arXiv180307058V}
{Valle}, G., {Dell'Omodarme}, M., {Prada Moroni}, P.~G., \& {Degl'Innocenti},
  S. 2018, ArXiv e-prints [\eprint[arXiv]{1803.07058}]

\bibitem[{{VandenBerg} {et~al.}(2006){VandenBerg}, {Bergbusch}, \&
  {Dowler}}]{2006ApJS..162..375V}
{VandenBerg}, D.~A., {Bergbusch}, P.~A., \& {Dowler}, P.~D. 2006, \apjs, 162,
  375

\bibitem[{{Zhang}(2013)}]{2013ApJS..205...18Z}
{Zhang}, Q.~S. 2013, \apjs, 205, 18

\end{thebibliography}
}

\end{document}